\documentclass[11pt]{article}

\usepackage[preprint]{acl}

\usepackage{times}
\usepackage{latexsym}

\usepackage[T1]{fontenc}

\usepackage[utf8]{inputenc}

\usepackage{microtype}

\usepackage{inconsolata}

\usepackage{graphicx}

\usepackage[utf8]{inputenc}
\usepackage[T1]{fontenc}

\usepackage{amsmath,amsfonts,amssymb}
\usepackage{graphicx}
\usepackage{booktabs}
\usepackage{multirow}
\usepackage{makecell}
\usepackage{array}
\usepackage{tabularx}
\usepackage{threeparttable}
\usepackage{colortbl}
\usepackage{xcolor}
\usepackage{subcaption}
\usepackage{caption}
\usepackage{adjustbox}
\usepackage{enumitem}
\usepackage{pifont}
\usepackage{ulem}
\usepackage{url}
\usepackage{microtype}
\usepackage{nicefrac}
\usepackage{siunitx}
\usepackage{graphicx}
\usepackage{arydshln}

\usepackage[ruled,vlined]{algorithm2e}

\usepackage{hyperref}
\hypersetup{hidelinks}

\title{LoSATok: Low-Dimensional Semantic-Acoustic Tokenizer for Cross-Domain Audio Understanding and Generation}

\author{
 \textbf{Zhisheng Zhang}\textsuperscript{1*\ensuremath{\dagger}},
 \textbf{Xiang Li}\textsuperscript{1\ensuremath{\dagger}},
 \textbf{Yixuan Zhou}\textsuperscript{1},
 \textbf{Jing Peng}\textsuperscript{1}, \\
 \textbf{Guoyang Zeng}\textsuperscript{2},
 \textbf{Zhiyong Wu}\textsuperscript{1\ensuremath{\ddagger}}
\\
 \textsuperscript{1}Shenzhen International Graduate School, Tsinghua University, China \\
 \textsuperscript{2}ModelBest Inc., China \\
 \texttt{zhangzs25@mails.tsinghua.edu.cn}
}

\begin{document}
\maketitle

\renewcommand{\thefootnote}{\fnsymbol{footnote}}
\footnotetext[1]{Work conducted while interning at ModelBest.}
\footnotetext[2]{Equal contribution.}
\footnotetext[3]{Corresponding author: Zhiyong Wu.}
\renewcommand{\thefootnote}{\arabic{footnote}}

\begin{abstract}
    Audio tokenizers are fundamental to unifying audio understanding and generation. Understanding requires high-level semantics, while generation demands semantic and acoustic details. Existing unified tokenizers jointly encode both in high-dimensional continuous latents, which increases the modeling burden of Diffusion Transformers (DiTs) for generation.
    We propose \textbf{LoSATok}, a low-dimensional audio tokenizer for cross-domain audio understanding and generation. Motivated by the observation that 1280-dimensional semantic encoder features are compressible, we introduce a Semantic Bottleneck that compresses them into 128 dimensions, regularized by the proposed time-relation loss for temporal feature consistency. We further design a dual-level semantic supervision method that leverages both high- and low-dimensional semantic signals, enabling the tokenizer to jointly capture semantics and acoustic details within a compact latent space.
    Experiments on speech, music, and general audio show that SemBo preserves strong low-dimensional semantic capacity and LoSATok retains competitive understanding performance compared with several semantic representations, while consistently improving DiT modeling performance on speech, music, and audio generation. 
    These results demonstrate that LoSATok's low-dimensional representations can effectively support audio understanding and generation.
    Our code is provided at \url{https://github.com/wxzyd123/LoSATok}.
\end{abstract}

\section{Introduction}
Understanding and generation are two important research directions in the audio domain~\cite{ming-uniaudio}. The former focuses on extracting high-level semantic information from audio signals to support tasks, e.g., automatic speech recognition (ASR) and audio event classification, while the latter requires models to produce natural and realistic audio signals, e.g., text-to-speech (TTS)~\cite{voxcpm}, text-to-music (TTM)~\cite{levo}, and text-to-audio (TTA)~\cite{uniflow-audio}. These two categories of tasks typically rely on distinct model architectures or representation designs. Although such separate designs have achieved promising results, they limit the model's ability to perform unified modeling across diverse tasks.

Recently, several works have explored joint modeling of understanding and generation, which can be categorized into model-level and representation-level approaches. At the \textbf{model level}, unified audio language models, e.g., UALM~\cite{ualm}, enable a single model to handle both understanding and generation tasks via post-training, reasoning, etc. Such methods generally require more complex training pipelines, and their internal representations may still fail to simultaneously satisfy the requirements of semantic preservation, acoustic reconstruction, and efficient modeling. 
Another line of work aims to unify understanding and generation at the \textbf{representation level}, i.e., through a unified audio representation. DashengTokenizer~\cite{dashengtokenizer} injects acoustic details into a semantic representation to jointly model semantics and acoustics, exhibiting stronger generality than purely acoustic representations across various understanding and generation tasks. 
RAE~\cite{rae} shows that semantically rich continuous representations can improve generative modeling with Diffusion Transformers (DiTs)~\cite{dit}. 
However, these high-dimensional representations often increase the modeling burden on downstream generators, requiring wider DiT blocks or more parameters to converge effectively.

In this paper, we consider a research question: {\it Can we learn a \uline{low-dimensional}, \uline{semantically rich}, and \uline{acoustically reconstructable} audio representation that supports \uline{cross-domain} modeling for both \uline{understanding and generation}?} 
Acoustic signals are usually low-dimensional~\cite{uniflow-audio}, but existing semantic representations are mostly high-dimensional, e.g., $>768$ dimensions. Therefore, a direct low-dimensional semantic signal is lacking. 
To this end, we investigate the intrinsic structure of the high-dimensional semantic representations from MiDashengLM~\cite{midashenglm}. We find that these semantic representations exhibit information redundancy: \textbf{their variance is not uniformly distributed across dimensions but instead concentrated in some principal directions}. Effective rank and Principal Component Analysis (PCA) reveal that a substantial portion of the semantic information is preserved even after compressing the high-dimensional representations to \textbf{128} dimensions. Training-free channel merging or PCA-based reduction retains rich semantics at 128 dimensions. However, these methods are not optimized for temporal structure, making them suboptimal as semantic supervision.

Building on this, we propose the \textbf{Sem}antic \textbf{Bo}ttleneck (\textbf{SemBo}) to learn a low-dimensional semantic signal from high-dimensional semantic representations. Specifically, we freeze a pre-trained semantic encoder and employ a compressor to project the high-dimensional representation into a low-dimensional space and a restorer to reconstruct it. 
In addition to the high-dimensional reconstruction loss, we expect the high-dimensional semantic signal to directly supervise the low-dimensional one, while the two have different dimensionalities. Inspired by the Gram loss~\cite{style-transfer}, we propose a time-relation loss that preserves the semantic relationships across different temporal frames. Compared with training-free strategies, SemBo explicitly models the relationship between low- and high-dimensional semantic representations via a learnable non-linear mapping, yielding a low-dimensional semantic representation that is better suited as a supervision signal.

Based on this low-dimensional semantic signal, we propose the \textbf{Lo}w-dimensional \textbf{S}emantic-\textbf{A}coustic \textbf{Tok}enizer (\textbf{LoSATok}), an architecture for cross-domain audio representation modeling. 
To jointly encode semantic and acoustic features, we introduce a dual-level semantic supervision scheme. The high-dimensional semantic supervision provides a complete semantic target, while the low-dimensional semantic supervision serves as a direct and compact target that guides the tokenizer to also learn low-dimensional semantic features. 
Meanwhile, we inject acoustic signals into the tokenizer, enabling native joint modeling of semantic and acoustic properties.

We comprehensively assess the effectiveness of LoSATok across speech, music, and general audio domains. 
Experiments across speech, music, and general audio show that SemBo retains approximately 93\% of the teacher encoder’s average understanding performance at 10\% of the feature dimension, even better than several Self-Supervised Learning (SSL) representations, e.g., HuBERT and WavLM. LoSATok provides a compact reconstructable latent whose sampled view preserves strong global and event-level understanding, while trading off fine-grained content probing for improved DiT learnability. Under a 208M-parameter DiT, LoSATok outperforms the acoustic baseline in both single-task and multi-task training, while reducing the DiT inference cost from 1,926 to 944 GFLOPs. These results position LoSATok as a generation-efficient semantic–acoustic tokenizer.
The main contributions are summarized as follows:
\begin{itemize}[leftmargin=1em, topsep=1pt, itemsep=1pt, parsep=1pt]
    \item We analyze the high-dimensional semantic representations' redundancy, reveal their low-rank structure, and verify the feasibility of low-dimensional semantic supervision.
    \item We propose the Semantic Bottleneck, which compresses high-dimensional semantic signals with a learnable network via semantic reconstruction and time-relation preservation.
    \item We propose LoSATok with dual-level semantic supervision, enabling a 128-dimensional reconstructable semantic–acoustic latent and explicitly characterizing its trade-offs across semantic probing, acoustic reconstruction, and generation.
    \item We validate the efficiency and effectiveness of the proposed low-dimensional unified representation across speech, music, and general audio, covering 15 understanding tasks and three generation tasks (TTA, TTM, and TTS).
\end{itemize}

\section{Related Work}

\noindent\textbf{Unified Understanding and Generation.}
Audio understanding and generation have previously relied on different models or representations. Some recent studies have explored unified modeling of the two. One line of work seeks unification at the model level. UALM~\cite{ualm} unifies understanding, generation, and reasoning within one model through different training strategies, such as SFT and DPO. Ming-UniAudio~\cite{ming-uniaudio} targets speech scenarios and uses high-dimensional general representations for LLM modeling and low-dimensional representations for DiT modeling, thereby unifying speech understanding, generation, and editing. These methods achieve model-level unification through more complex training pipelines and engineering architectures. 
Another line of work operates at the representation level, aiming to support both understanding and generation within a single general-purpose representation. 
JMAS-VAE~\cite{jmas-vae} utilizes joint-marginal distillation losses and adaptive weighting, improving the balance between speech reconstruction, understanding, and generation.
DashengTokenizer~\cite{dashengtokenizer} injects acoustic information into frozen semantic features to construct a continuous audio tokenizer, enabling joint semantic-acoustic modeling. However, such high-dimensional representations increase the burden on downstream DiT modeling and require wider DiT blocks. 
Concurrently, WavCube~\cite{wavcube} models speech understanding and generation by directly compressing high-dimensional semantic representations into 128-dimensional semantic representations and constructing acoustic details from them. Our work differs in four respects: 1) we jointly and natively model semantics and acoustics, reducing the difficulty of modeling acoustics from semantic representations; 2) we analyze the compressibility of high-dimensional semantic representations and propose a time-relation loss to explicitly learn the mapping between high- and low-dimensional representations; 3) we achieve cross-domain understanding and generation beyond the speech domain; and 4) we investigate the efficiency and performance of acoustic and semantic tokenizers for downstream DiT-based generation.

\noindent\textbf{Discrete and Continuous Audio Tokenizers.} 
Audio tokenizers have been widely used across audio-related tasks, e.g., ASR and TTS. Neural audio codecs (NACs) typically use discrete token representations~\cite{unisrcodec}. DAC~\cite{dac} and SNAC~\cite{snac} quantize encoder representations to achieve efficient audio compression. Existing discrete NACs can be easily combined with autoregressive language models~\cite{llasa}, but they suffer from quantization loss~\cite{voxcpm}, which limits their ability to perform fine-grained reconstruction.
Unlike discrete tokenizers, continuous audio tokenizers (CATs) model audio directly in the continuous latent space. They usually offer stronger reconstruction capabilities and are better suited to continuous generative architectures such as DiT. The UniFlow-Audio~\cite{uniflow-audio} architecture, based on an acoustic VAE, achieves superior reconstruction. Recent work introduces semantic information into continuous tokenizers to bridge the gap between acoustic and semantic representations. SemanticVocoder~\cite{semanticvocoder} reconstructs audio from semantic latents and restores acoustic details from semantic information using a generative decoder.
This paper focuses on CATs, aiming to preserve more acoustic details with rich semantics and better adapt to DiTs.

\section{Semantic Bottleneck}\label{sec:sembo}

\begin{figure}[t]
    \centerline{
    \includegraphics[width=0.45\textwidth]{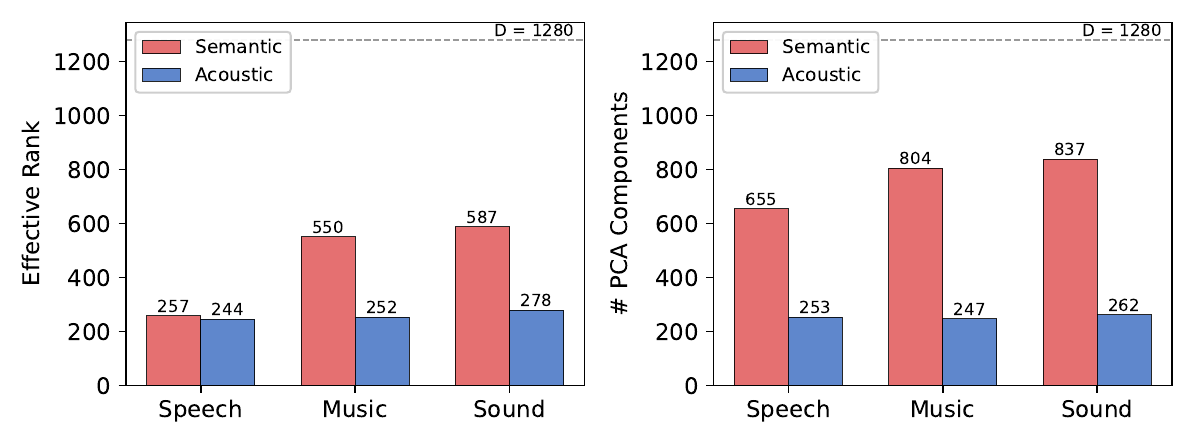}}
    \caption{Effective rank and PCA components.}
    \label{fig:dim}
\end{figure}

\noindent\textbf{Effective Rank and PCA Analysis.}
To obtain low-dimensional semantic supervision, we explore whether high-dimensional semantic representations are compressible. 
If semantic representations concentrate along a few principal directions, they contain redundancy. We analyze acoustic and semantic representations of DashengTokenizer using effective rank and Principal Component Analysis (PCA). 
The semantic representations are extracted from MiDashengLM~\cite{midashenglm}.

\begin{figure*}[t]
    \centerline{
    \includegraphics[width=0.9\textwidth]{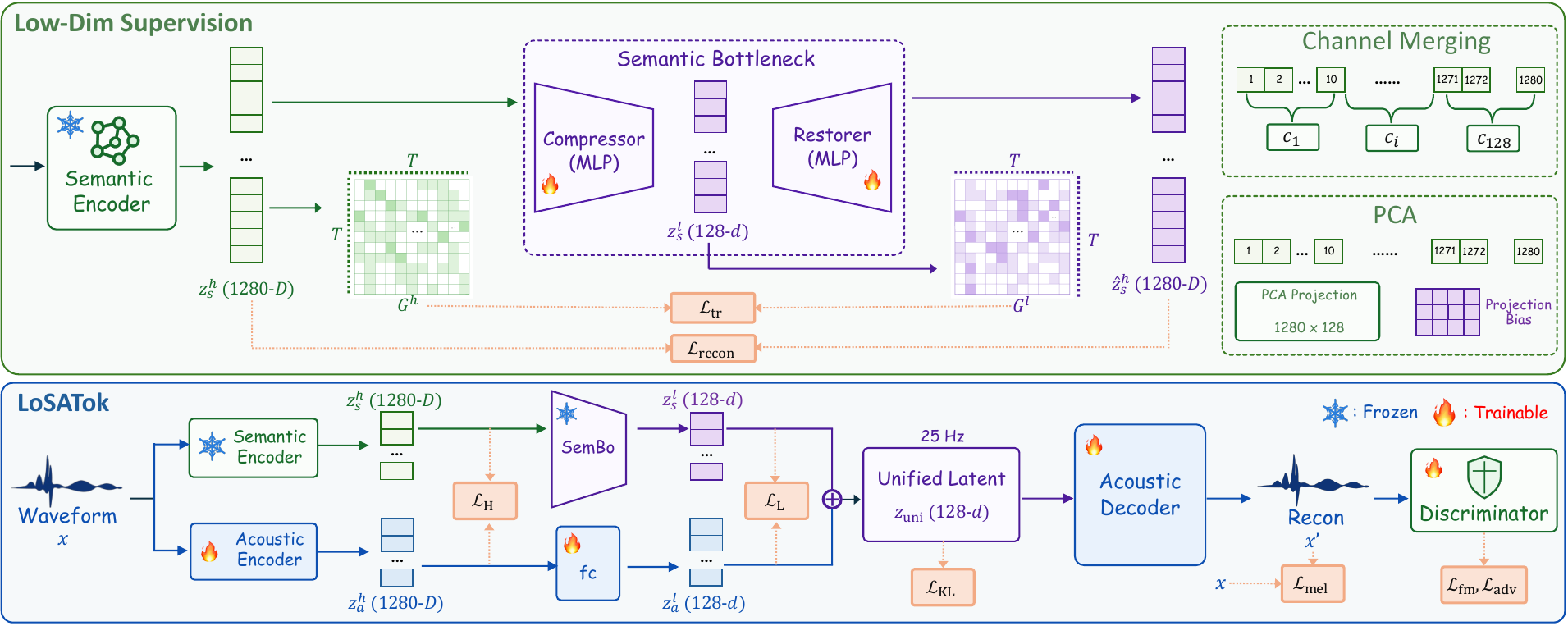}}
    \caption{The architecture and training components of SemBo and LoSATok.}
    \label{fig_workflow}
\end{figure*}

As shown in Figure \ref{fig:dim}, although the original feature dimension is 1280, the effective rank is only 257 on speech and remains below 50\% of the original dimension in the music and audio domains of semantic representations, suggesting that semantic energy concentrates along dominant directions. 
PCA also shows that fewer than 1280 components are sufficient to preserve 90\% of the variance. 
These results indicate that semantic representations are more complex than acoustic features but still contain substantial dimensional redundancy, which motivates subsequent compression. Further details are provided in Appendix \ref{sec:details_rank}.

\noindent\textbf{Training-Free Dimensionality Reduction.}
Based on the above low-rank property, we test whether training-free dimensionality reduction preserves semantic information. 
The first method is \textit{channel merging} (CM). For a 1280-dimensional semantic representation, we average every 10 channels and obtain a 128-dimensional representation. 
The second method is \textit{PCA reduction}. We randomly sample 6000 examples from the cross-domain data in Table \ref{tab:datasets}, fit PCA using extracted semantic features, and retain the first 128 principal components.

\begin{table}[t]
    \centering
    \caption{Understanding on the XARES benchmark.}
    \label{tab:dim_reduction}
    \resizebox{0.4\textwidth}{!}{
    \begin{tabular}{lcccc}
    \toprule
    \textbf{Method} 
    & \textbf{Dimension} 
    & \textbf{ESC}($\uparrow$) 
    & \textbf{FSC}($\uparrow$) 
    & \textbf{GTZAN}($\uparrow$) \\
    \midrule
    MiDashengLM 
        & 1280 & 96.95 & 98.26 & 91.19 \\
    \hdashline
    CM 
        & 128 & 92.80 & 86.11 & 89.39 \\
    PCA 
        & 128 & \textbf{94.95} & 78.06 & 90.49 \\
    \rowcolor{gray!15} \textbf{SemBo (Ours)} 
        & 128 & 93.75 & \textbf{89.11} & \textbf{89.29} \\
    \bottomrule
    \end{tabular}
    }
\end{table}

Table \ref{tab:dim_reduction} shows that both methods preserve semantic information on downstream understanding tasks, confirming that the semantic representation has a compressible redundant structure. 
However, these methods do not explicitly model the mapping between high- and low-dimensional semantic latents, and thus provide weak supervision for unified representations, as discussed in Section \ref{sec:exp_ablations}.

\noindent\textbf{Semantic Bottleneck.}
To obtain a more effective low-dimensional semantic signal, we propose \textbf{Sem}antic \textbf{Bo}ttleneck (\textbf{SemBo}). SemBo consists of a compressor and a restorer, as shown in Figure \ref{fig_workflow}. Given the high-dimensional output $z^{h}_s$ from a frozen semantic encoder, the compressor $C$ maps it to $z^{l}_s = C(z^h_s), \quad z^l_s \in \mathbb{R}^{T \times d}$,
where $d=128$. The restorer $R$ then reconstructs the high-dimensional representation $\hat{z}^{h}_s = R(z^{l}_s).$

Both modules are implemented with lightweight 2-layer multilayer perceptrons (MLPs). The reconstruction loss is given in Eq. (\ref{eq:sembo_recon}).
\begin{equation}
    \mathcal{L}_{\mathrm{recon}} = \left\| \text{norm}\left( \hat{z}^{h}_s \right) - \mathrm{sg}\left(\text{norm}(z^{h}_s)  \right) \right\|_2,
    \label{eq:sembo_recon}
\end{equation}
where $\mathrm{sg}(\cdot)$ denotes stop-gradient and $\text{norm}(\cdot)$ denotes normalization.
Since reconstruction alone does not directly constrain the low-dimensional representation, we introduce a time-relation loss to achieve direct high-dimensional supervision for $z^l_s$. First, we compute temporal similarity matrices with input representation normalization
$\mathbf{G}^{h} = z^{h}_s (z^{h}_s)^{\top} \in \mathbb{R}^{T \times T}, \mathbf{G}^{l} = z^{l}_s (z^{l}_s)^{\top} \in \mathbb{R}^{T \times T}$ for high- and low-dimensional representations. Then, we align the temporal relation of $\mathbf{G}^{h}$ and $\mathbf{G}^{l}$:
\begin{equation}
    \mathcal{L}_{\mathrm{tr}} = \left\| \mathbf{G}^{l} - \mathrm{sg}(\mathbf{G}^{h}) \right\|_2.
    \label{eq:sembo_tr}
\end{equation}

\noindent\textbf{Intuition.} Direct feature-wise alignment between $z_s^h$ and $z_s^l$ is infeasible because they have different channel dimensions. However, their normalized temporal Gram matrices are both $T\times T$ and describe the same dimension-independent temporal geometry: which frames are semantically similar and which are dissimilar. Matching these matrices therefore transfers the temporal-semantic relations of the teacher space into the bottleneck. Moreover, $\mathcal{L}_{\mathrm{recon}}$ constrains only the composite mapping $R\circ C$, leaving the bottleneck under-determined under many invertible reparameterizations. In contrast, $\mathcal{L}_{\mathrm{tr}}$ directly anchors the geometry of $z_s^l$, explaining why removing it causes the larger downstream degradation in Fig. \ref{fig:ablation_sembo}.

The final objective for SemBo is in Eq. (\ref{eq:sembo}).
\begin{equation}
    \mathcal{L}_{\mathrm{SemBo}} = \lambda_{\mathrm{recon}} \mathcal{L}_{\mathrm{recon}} + \mathcal{L}_{\mathrm{tr}},
    \label{eq:sembo}
\end{equation}
where $\lambda_{\mathrm{recon}}=10^3$ with hyperparameter sensitivity analysis in Section \ref{sec:exp_ablations}.
Compared with CM and PCA, SemBo performs well on FSC and GTZAN tasks in Table \ref{tab:dim_reduction} and explicitly learns a nonlinear mapping between high- and low-dimensional semantic spaces, making it better suited as the low-dimensional semantic supervision for LoSATok.

\section{LoSATok}

\noindent\textbf{Problem Definition.}
Given an input waveform $x$, LoSATok maps it with an encoder into a continuous latent representation $z_{\mathrm{uni}} \in \mathbb{R}^{T \times d}$, where $d$ denotes the latent dimension and is set to $128$. A unified (i.e., semantic-acoustic) representation should:
\begin{enumerate}[leftmargin=1em, topsep=0pt, itemsep=0pt, parsep=0pt]
    \item \textit{Semantic Richness:} contain rich semantic information for downstream understanding models.
    \item \textit{Acoustic Reconstruction:} preserve basic acoustic details for reconstruction.
    \item \textit{Generation Efficiency:} provide a compact and effective latent space for generative models.
\end{enumerate}
Therefore, the objective of LoSATok is to learn a low-dimensional semantic-acoustic representation.

\noindent\textbf{Dual-Level Semantic Supervision.}
Let the high-dimensional acoustic representation be $z_a^{h} = \mathrm{E}_a(x) \in \mathbb{R}^{T \times 1280}$ and the low-dimensional acoustic representation be $z_a^{l} = \mathrm{fc}\left(\mathrm{E}_a(x)\right)$, where $\mathrm{E}_a$ is the acoustic encoder that encodes the waveform into a high-dimensional acoustic representation and $\text{fc}(\cdot)$ represents a linear layer for acoustic channel compression. 
For semantic-acoustic joint modeling, we aim to obtain the unified representation $z_{\mathrm{pre}} = z^l_a + z^l_s$. Then, we obtain $\mu,\log\sigma^2=f_{\rm KL}(z_{\rm pre}),\quad z_{\rm uni}=\mu+\sigma\odot\epsilon$.

To enable $z_{\mathrm{uni}}$ to capture both semantic and acoustic features, we introduce dual-level semantic supervision. Specifically, we use both high- and low-dimensional semantic signals to supervise the latent space. The high-dimensional semantic representation provides global semantic information, while the low-dimensional semantic representation directly constrains the acoustic features. These two levels complement each other, allowing LoSATok to preserve generation-friendly representations:
\begin{equation}
    \mathcal{L}_{\mathrm{H}} = \left\| z_a^{h} - \mathrm{sg}(z_s^{h}) \right\|_2, \ 
    \mathcal{L}_{\mathrm{L}} = \left\| z_a^{l} - \mathrm{sg}(z_s^{l}) \right\|_2
\end{equation}

\noindent\textbf{Architecture.} LoSATok utilizes an encoder-decoder architecture. The encoder consists of a semantic encoder and an acoustic encoder. For the semantic encoder, we use the MiDashengLM~\cite{midashenglm} audio encoder, which is pretrained on large-scale data. For the acoustic encoder, following \cite{dashengtokenizer}, we adopt a non-overlapping patch embedding scheme to align the acoustic and semantic representations. The decoder is based on Vocos~\cite{vocos} and generates audio from the unified representation. The frame rate of LoSATok is 25Hz, and detailed architecture information is shown in Appendix \ref{sec:details_losatok}.

\noindent\textbf{Training Objectives.}
In addition to semantic capabilities, LoSATok must also ensure audio reconstruction capability. Following~\cite{dac}, we improve the naturalness and fine-grained quality of reconstructed audio using a mel reconstruction loss $\mathcal{L}_{\mathrm{mel}}$ and adversarial training. 
We adopt a Multi-Frequency Discriminator (MFD)~\cite{llasa} with a hinge loss objective. The training objective of LoSATok is given in Eq. (\ref{eq:losatok}).
\begin{equation}
\begin{aligned}
    \mathcal{L} 
    = \lambda_{\mathrm{mel}} \mathcal{L}_{\mathrm{mel}}
    + \lambda_{\mathrm{sem}}
    \left(\mathcal{L}_{\mathrm{H}} + \mathcal{L}_{\mathrm{L}}\right) + \\
    \lambda_{\mathrm{KL}} \mathcal{L}_{\mathrm{KL}}
    + \lambda_{\mathrm{fm}} \mathcal{L}_{\mathrm{fm}}
    + \lambda_{\mathrm{adv}} \mathcal{L}_{\mathrm{adv}},
\end{aligned}
\label{eq:losatok}
\end{equation}
where $\mathcal{L}_{\mathrm{fm}}$ is the feature-matching loss, and $\mathcal{L}_{\mathrm{adv}}$ is the adversarial loss. $\mathcal{L}_{\mathrm{KL}}$ is the Kullback-Leibler (KL) divergence loss. The loss weights are set to $\{\lambda_{\mathrm{mel}}, \lambda_{\mathrm{sem}}, \lambda_{\mathrm{KL}}, \lambda_{\mathrm{fm}}, \lambda_{\mathrm{adv}} \} = \{45, 45, 10^{-2}, 1, 1\}$. 
Details of these objectives are provided in Appendix \ref{sec:details_losatok}. 

\noindent\textbf{Downstream DiT Generation.}
The unified representation $z_{\mathrm{uni}}$ can serve as the target representation for downstream generative models. For TTA, TTM, and TTS tasks, given a text condition $c$, DiT learns to generate LoSATok latents from noise, modeled as $p_{\theta}(z_{\mathrm{uni}} \mid c)$. The generated latents are then converted back into audio by the LoSATok decoder.

\section{Experiments}

\subsection{LoSATok Training Setup}
\noindent\textbf{Datasets.} We select a diverse 13.2K-hour cross-domain dataset, consisting of 34.6\% speech data (LibriSpeech~\cite{librispeech}, VCTK~\cite{vctk}, and the Common Voice-en subset~\cite{common_voice}), 28.6\% music data (MTG-Jamendo~\cite{jamendo} and MUSDB~\cite{musdb18}), and 36.8\% general audio data (AudioSet~\cite{audioset}).

\noindent\textbf{Training Details.} We train the model for one million steps on 8 NVIDIA H100 GPUs. The initial learning rate is set to $1 \times 10^{-4}$, and optimization is performed using AdamW. The global batch size is set to 64 during training.

\subsection{Audio Understanding}
\begin{table*}[t]
    \centering
    \caption{Understanding evaluation of representations on speech, music, and general audio domains. ``Latent Dim.'' denotes the representation dimension of tokenizers, and ``Avg.'' denotes the average scores of understanding tasks.}
    \scriptsize
    \resizebox{\textwidth}{!}{
    \begin{tabular}{l c c c c c c c c c c c c c c c c c}
    \toprule
    \multirow{2}{*}{\textbf{Model}} 
    & \multirow{2}{*}{\textbf{\makecell[c]{Latent \\ Dim.}}} 
    & \multicolumn{7}{c}{\textbf{Speech Understanding}($\uparrow$)} 
    & \multicolumn{4}{c}{\textbf{Music Understanding}($\uparrow$)} 
    & \multicolumn{4}{c}{\textbf{Audio Understanding}($\uparrow$)}
    & \multirow{2}{*}{\textbf{Avg.}}  \\
    \cmidrule(lr){3-9} \cmidrule(lr){10-13} \cmidrule(lr){14-17}
    & & LS100h & CD & FSC & LibCnt & LSMF & RAV & VocS 
    & FMA & GTZAN & MT & NSynth 
    & Clo & DES & ESC & Urb8 & \\
    \midrule
    \multicolumn{18}{l}{\textit{Low-dimensional Tokenizers}} \\
    EnCodec         
        & 128  
        & 0.00  & 39.16 & 3.48  & 33.78 & 88.44 & 27.71 & 43.06 
        & 28.80 & 35.14 & 23.37 & 32.59 & 0.57 & 3.82  & 16.80 & 40.34
        & 27.80 \\
    EzAudio         
        & 64   
        & 0.00  & 38.80 & 1.63  & 37.19 & 84.90 & 25.56 & 35.26 
        & 30.74 & 40.54 & 17.57 & 32.18 & 0.40 & 17.01 & 18.05 & 34.97
        & 27.65 \\
    UniFlow-Audio   
        & 128  
        & 0.00  & 39.78 & 1.32  & 37.15 & 74.57 & 23.26 & 34.73 
        & 30.17 & 40.74 & 16.44 & 31.73 & 0.57 & 16.90 & 19.85 & 35.16
        & 26.82 \\
    \rowcolor{blue!5} \textbf{LoSATok} & & & & & & & & & & & & & & & & & \\

    \rowcolor{blue!5} \quad \quad $z_{s}^l$
        & \textcolor{red}{128} 
        & 89.16 & 72.85 & 89.11 & 75.28 & 96.86 & 71.25 & 90.59 
        & 65.37 & 89.29 & 22.86 & 63.16 & 6.11 & 46.72 & 93.75 & 84.94
        & \textbf{70.49} \\
    \rowcolor{blue!5} \quad \quad $z_{\mathrm{pre}}$
        & \textcolor{red}{128} 
        & 82.74 & 70.34 & 57.13 & 71.66 & 97.08 & 59.79 & 90.18
        & 63.43 & 87.09 & 44.25 & 59.62 & 5.46 & 38.47 & 90.95 & 84.47 
        & \uline{66.84} \\
    \rowcolor{blue!5} \quad \quad $\mu$
        & \textcolor{red}{128}  
        & 61.67 & 67.65 & 32.48 & 68.01 & 95.51 & 65.28 & 88.51
        & 59.54 & 85.89 & 44.27 & 51.76 & 4.78 & 39.75 & 89.55 & 83.04
        & 62.51 \\
    \rowcolor{blue!5} \quad \quad $z_{\mathrm{uni}}$
        & \textcolor{red}{128}  
        & 39.00 & 65.41 & 25.10 & 66.63 & 95.14 & 60.42 & 88.03
        & 58.06 & 85.39 & 41.87 & 51.39 & 4.86 & 36.67 & 88.90 & 82.66
        & 59.30 \\

    \midrule
    \multicolumn{18}{l}{\textit{Generalization of SemBo}} \\
    Qwen3-Omni
        & 2048
        & 85.03 & 67.65 & 95.54 & 83.39 & 98.80 & 59.93
        & 86.86 & 55.66 & 78.68 & 17.68
        & 32.67 & 1.76 & 32.67 & 74.85
        & 77.29 & 63.23 \\

    \quad\quad \textbf{w/ SemBo}
        & \textcolor{red}{128}
        & 83.79 & 61.65 & 86.16 & 70.19 & 98.92 & 43.06
        & 77.76 & 57.71 & 70.07 & 13.49
        & 33.69 & 1.73 & 27.61 & 63.00
        & 73.79 & 57.51 \\
    
    \midrule
    \multicolumn{18}{l}{\textcolor{gray}{\textit{High-dimensional Tokenizers}}} \\
    \textcolor{gray}{DAC}             
        & \textcolor{gray}{1024} 
        & \textcolor{gray}{0.00}  & \textcolor{gray}{43.90} & \textcolor{gray}{2.55}  & \textcolor{gray}{43.93} & \textcolor{gray}{91.58} & \textcolor{gray}{36.45} & \textcolor{gray}{42.17} 
        & \textcolor{gray}{35.43} & \textcolor{gray}{54.06} & \textcolor{gray}{12.78} & \textcolor{gray}{38.57} & \textcolor{gray}{0.73} & \textcolor{gray}{20.20} & \textcolor{gray}{31.20} & \textcolor{gray}{50.26}
        & \textcolor{gray}{33.59}\\
    \textcolor{gray}{SemantiCodec}   
        & \textcolor{gray}{768}  
        & \textcolor{gray}{0.00}  & \textcolor{gray}{60.30} & \textcolor{gray}{46.40} & \textcolor{gray}{66.05} & \textcolor{gray}{98.14} & \textcolor{gray}{55.00} & \textcolor{gray}{84.28} 
        & \textcolor{gray}{57.94} & \textcolor{gray}{66.67} & \textcolor{gray}{8.62}  & \textcolor{gray}{68.14} & \textcolor{gray}{1.89} & \textcolor{gray}{48.57} & \textcolor{gray}{81.75} & \textcolor{gray}{84.54}
        & \textcolor{gray}{55.22} \\
    \textcolor{gray}{Whisper}         
        & \textcolor{gray}{1280} 
        & \textcolor{gray}{90.00} & \textcolor{gray}{71.32} & \textcolor{gray}{97.78} & \textcolor{gray}{64.40} & \textcolor{gray}{94.93} & \textcolor{gray}{68.47} & \textcolor{gray}{91.48} 
        & \textcolor{gray}{58.90} & \textcolor{gray}{71.77} & \textcolor{gray}{0.00}  & \textcolor{gray}{63.50} & \textcolor{gray}{3.10} & \textcolor{gray}{22.64} & \textcolor{gray}{62.45} & \textcolor{gray}{75.74}
        & \textcolor{gray}{62.43} \\
    \textcolor{gray}{HuBERT}          
        & \textcolor{gray}{1024} 
        & \textcolor{gray}{82.45} & \textcolor{gray}{58.06} & \textcolor{gray}{98.73} & \textcolor{gray}{49.72} & \textcolor{gray}{85.94} & \textcolor{gray}{47.71} & \textcolor{gray}{81.71} 
        & \textcolor{gray}{42.81} & \textcolor{gray}{51.85} & \textcolor{gray}{0.00}  & \textcolor{gray}{48.22} & \textcolor{gray}{1.74} & \textcolor{gray}{3.64}  & \textcolor{gray}{37.35} & \textcolor{gray}{57.40}
        & \textcolor{gray}{49.82} \\
    \textcolor{gray}{WavLM}           
        & \textcolor{gray}{1024} 
        & \textcolor{gray}{67.06} & \textcolor{gray}{45.16} & \textcolor{gray}{96.07} & \textcolor{gray}{50.37} & \textcolor{gray}{75.95} & \textcolor{gray}{32.57} & \textcolor{gray}{71.19} 
        & \textcolor{gray}{36.11} & \textcolor{gray}{48.14} & \textcolor{gray}{0.00}  & \textcolor{gray}{40.14} & \textcolor{gray}{0.79} & \textcolor{gray}{16.76} & \textcolor{gray}{31.45} & \textcolor{gray}{53.13}
        & \textcolor{gray}{44.33} \\
    \textcolor{gray}{MingTok-Audio}   
        & \textcolor{gray}{896}  
        & \textcolor{gray}{93.45} & \textcolor{gray}{66.85} & \textcolor{gray}{98.58} & \textcolor{gray}{62.92} & \textcolor{gray}{97.10} & \textcolor{gray}{57.08} & \textcolor{gray}{88.59} 
        & \textcolor{gray}{54.45} & \textcolor{gray}{71.17} & \textcolor{gray}{25.00} & \textcolor{gray}{57.00} & \textcolor{gray}{2.00} & \textcolor{gray}{39.80} & \textcolor{gray}{62.25} & \textcolor{gray}{72.87}
        & \textcolor{gray}{63.27} \\
    \textcolor{gray}{DashengTok} 
        & \textcolor{gray}{1280} 
        & \textcolor{gray}{68.71} & \textcolor{gray}{80.56} & \textcolor{gray}{83.39} & \textcolor{gray}{79.98} & \textcolor{gray}{98.06} & \textcolor{gray}{83.06} & \textcolor{gray}{93.15} 
        & \textcolor{gray}{66.29} & \textcolor{gray}{89.99} & \textcolor{gray}{57.65} & \textcolor{gray}{77.91} & \textcolor{gray}{5.64} & \textcolor{gray}{55.40} & \textcolor{gray}{96.40} & \textcolor{gray}{83.80}
        & \textcolor{gray}{74.67} \\
    \textcolor{gray}{MiDashengLM}        
        & \textcolor{gray}{1280} 
        & \textcolor{gray}{80.32} & \textcolor{gray}{80.10} & \textcolor{gray}{98.26} & \textcolor{gray}{80.79} & \textcolor{gray}{98.47} & \textcolor{gray}{87.15} & \textcolor{gray}{93.15} 
        & \textcolor{gray}{65.14} & \textcolor{gray}{91.19} & \textcolor{gray}{38.65} & \textcolor{gray}{77.59} & \textcolor{gray}{6.12} & \textcolor{gray}{52.02} & \textcolor{gray}{96.95} & \textcolor{gray}{86.23}
        & \textcolor{gray}{75.48} \\
    \bottomrule
    \end{tabular}
    }
    \label{tab:audio_understanding}
\end{table*}

\noindent\textbf{Evaluation Setup.} For understanding evaluation, we use the XARES benchmark~\cite{xares} with a linear probing protocol across 15 cross-domain tasks. XARES normalizes all task results to a $0 \sim 1$ scale, with higher values indicating better performance. We report percentages (\%).

\noindent\textbf{Speech Understanding.}
We evaluate 7 speech understanding tasks: ASR (LS100h), emotion recognition (Crema-D, RAV), intent classification (FSC), speaker counting (LibCnt), gender classification (LSMF), and vocal sound classification (VocS). For ASR, we connect the representation to Qwen2.5-0.5B for evaluation, while for the other tasks, we use the MLP classifier as in XARES.

\noindent\textbf{Music Understanding.}
We evaluate the representation capability on music understanding tasks, including music genre classification (FMA, GTZAN), music transcription (MAESTRO, MT), and musical timbre/source recognition (Nsynth).

\noindent\textbf{Audio Understanding.}
We select 4 types of general audio understanding tasks: audio captioning (Clo), sound/environment classification (ESC, Urb8), and domestic sound event detection (DES).

\noindent\textbf{Results.}
Table~\ref{tab:audio_understanding} reports the understanding performance of the different LoSATok views. The 128-dimensional semantic view $z_s^l$ achieves an average score of 70.49 across 15 tasks, retaining 93.4\% of the performance of the 1280-dimensional MiDashengLM (75.48). Following the XARES ASR protocol, which connects the representation to Qwen2.5-0.5B, $z_s^l$ obtains 89.16 on LS100h, approaching the performance of the ASR-oriented Whisper representation.
The deterministic pre-KL unified embedding $z_{\mathrm{pre}}$ achieves an average score of 66.84, including 82.74 on LS100h, indicating that substantial linguistic information remains accessible after acoustic-detail injection. The posterior mean $\mu$ obtains 62.51 on average, while the reparameterized latent $z_{\mathrm{uni}}$ obtains 59.30, still outperforming several high-dimensional SSL representations, including \textit{HuBERT~\cite{hubert} and WavLM~\cite{wavlm}}. The effect of KL regularization and sampling is concentrated on fine-grained content tasks: from $z_{\mathrm{pre}}$ to $z_{\mathrm{uni}}$, LS100h decreases from 82.74 to 39.00 and FSC from 57.13 to 25.10. In contrast, global and event-level tasks are considerably more stable, with LSMF changing from 97.08 to 95.14 and ESC-50 from 90.95 to 88.90. Moreover, acoustic-detail injection benefits tasks whose relevant information is primarily acoustic: on MT, the unified latent improves the score from 22.86 to 41.87 over the semantic view. Overall, these results reveal a structured trade-off: deterministic views provide stronger accessibility for fine-grained understanding, whereas the sampled unified latent retains competitive broad-domain understanding while offering the regularized, reconstructable representation required for downstream generation.

\noindent\textbf{Generalization of SemBo on Qwen3-Omni.} To test whether the observed redundancy is specific to MiDashengLM, we adapt the analysis using the 2048-dimensional Qwen3-Omni~\cite{qwen3-omni} audio encoder. Its effective rank is only approximately 100$\sim$125 across domains, and PCA-128 already explains 84$\sim$90\% of the variance. We further train SemBo to compress the Qwen3-Omni features from 2048 to 128 dimensions, corresponding to a $16\times$ reduction. As shown in Table \ref{tab:audio_understanding}, SemBo retains 57.51/63.23, or approximately 91\%, of the average understanding score, while preserving ASR performance (83.79 vs. 85.03). These results suggest that both the redundancy pattern and semantic bottleneck generalize beyond the MiDashengLM.

\begin{table*}[t]
    \centering
    \caption{Downstream generation results of \textit{general-purpose} continuous audio tokenizers (CATs). ``GFLOPs'' represents the inference efficiency of DiT sampling process. The best results are highlighted in \textbf{bold}.}
    \resizebox{\textwidth}{!}{
    \begin{tabular}{lcccccccccccccc}
    \toprule
    \multirow{2}{*}{\textbf{Model}} 
    & \multirow{2}{*}{\textbf{\makecell{Latent \\ Dim.}}} 
    & \multirow{2}{*}{\textbf{\makecell{DiT \\ Dim.}}}
    & \multirow{2}{*}{\textbf{GFLOPs}($\downarrow$)}
    & \multicolumn{3}{c}{\textbf{Text-To-Speech}} 
    & \multicolumn{4}{c}{\textbf{Text-To-Music}} 
    & \multicolumn{4}{c}{\textbf{Text-To-Audio}} \\
    \cmidrule(lr){5-7} \cmidrule(lr){8-11} \cmidrule(lr){12-15}
    & & & 
    & WER($\downarrow$) & SIM($\uparrow$) & UTMOS($\uparrow$)
    & FAD($\downarrow$) & FD($\downarrow$) & KL($\downarrow$) & CLAP($\uparrow$)
    & FAD($\downarrow$) & FD($\downarrow$) & KL($\downarrow$) & CLAP($\uparrow$) \\
        \midrule
    \multicolumn{15}{c}{\textit{Single-task Training}} \\
    \midrule
    UniFlow-Audio 
    & 128 & 512 & 1926
    & 3.589 & 0.408 & 2.768
    & 6.147 & 36.098 & 1.807 & 0.250 
    & 4.925 & 40.017 & 2.613 & 0.243 \\

    \hdashline
    \multirow{3}{*}{DashengTok} 
    & 1280 & 512 & 1028
    & 100.0 & 0.015 & 1.251
    & 23.257 & 131.417 & 5.068 & 0.027 
    & 34.681 & 115.557 & 4.109 & 0.002 \\
    
    & 1280 & 1536 & 1910
    & 75.469 & 0.103 & 1.322
    & 8.460 & 51.246 & 1.832 & 0.237 
    & 7.238 & 43.916 & 2.395 & 0.245 \\
    
    & 1280 & 1536 & 7855
    & 3.652 & 0.287 & 3.144
    & \textbf{3.780} & 25.848 & \textbf{1.751} & 0.268
    & 4.138 & \textbf{24.605} & \textbf{1.738} & 0.379 \\

    \hdashline
    \rowcolor{blue!5} \textbf{LoSATok}
        & 128 & 512 & \textbf{944}
            & \textbf{3.030} & \textbf{0.548} & \textbf{3.367}
            & 4.156 & \textbf{25.089} & 1.788 & \textbf{0.282}
            & \textbf{2.760} & 25.743 & 1.844 & \textbf{0.381} \\

    \midrule
    \multicolumn{15}{c}{\textit{Multi-task Joint Training}} \\
    \midrule
    UniFlow-Audio
    & 128 & 512 & 1926
    & 4.529 & 0.344 & 2.259
    & 4.752 & 28.173 & 1.724 & 0.272 
    & 4.066 & 37.217 & 2.554 & 0.252 \\
    
    DashengTok
    & 1280 & 1536 & 1910 
    & 84.761 & 0.074 & 1.296
    & 5.780 & 36.227 & 1.797 & 0.268 
    & 4.482 & 37.146 & 2.137 & 0.283 \\
    \hdashline
    \rowcolor{blue!5} \textbf{LoSATok}
        & 128 & 512 & \textbf{944}
        & \textbf{3.667} & \textbf{0.507} & \textbf{3.310}
        & \textbf{3.366} & \textbf{20.127} & \textbf{1.628} & \textbf{0.300}
        & \textbf{1.987} & \textbf{21.363} & \textbf{1.700} & \textbf{0.396} \\
    \bottomrule
    \end{tabular}
    }
    \label{tab:generation}
\end{table*}

\subsection{Audio Generation}\label{sec:exp_generation}

\noindent\textbf{Evaluation Pipelines.}
To evaluate the downstream generation performance of LoSATok on TTA, TTM, and TTS tasks, we follow the evaluation protocol of DashengTokenizer and adopt UniFlow-Audio~\cite{uniflow-audio}, a flow-based DiT paradigm for cross-domain generation. Specifically, we replace the VAE in UniFlow-Audio with the tokenizer under evaluation. During training, the tokenizer is frozen, and the DiT architecture is trained, while other settings remain unchanged for a fair comparison.

In the main experiments, we use the same version of the generation architecture, where the hidden dimension of DiT is set to 512 and the number of layers is set to 12. This standard trainable DiT configuration contains approximately 208M parameters. For DashengTokenizer, since its representation dimension is 1280, we follow its original settings~\cite{dashengtokenizer} and consider three configurations to support training for comparison:
\begin{enumerate}[leftmargin=1em, topsep=0pt, itemsep=0pt, parsep=0pt]
    \item \textit{Align with the original VAE setting:} DiT dimension is 512, the layer number is 12, and the number of trainable parameters is 215M.
    \item \textit{Wider DiT block with comparable parameter:} DiT dimension is 1536, the layer number is 2, and the number of trainable parameters is 322M.
    \item \textit{Larger trainable parameter scale:} DiT dimension is 1536, the layer number is 12, and the number of trainable parameters is 975M.
\end{enumerate}

Moreover, the GFLOPs denote the inference efficiency of DiT. We evaluate this metric on LibriTTS with an average duration of 5.647 s per utterance.

\noindent\textbf{Training and Inference.}
During training, we consider both single-task training and multi-task joint training. For single-task training, all experiments are conducted on 4 NVIDIA H100 GPUs. Specifically, TTA is trained on the WavCaps~\cite{wavcaps} dataset for 100K steps, TTM is trained on the LP-MusicCaps-MTT~\cite{lp-musiccaps} subset for 50K steps, and TTS is trained on the LibriTTS~\cite{libritts} training set for 50K steps. For multi-task joint training, we merge the TTA, TTM, and TTS training data based on UniFlow-Audio, constructing a larger joint training setting with a more complex task composition, and train the model on 8 NVIDIA GPUs for 150K steps. In all training settings, the batch size on each GPU is set to 24.

During inference, the classifier-free guidance scale is 3.0, and the number of inference steps is 20.
During evaluation, we evaluate the models on the AudioCaps~\cite{audiocaps} test set, the MusicCaps~\cite{musiclm} test set, and the LibriTTS test set, respectively. For TTA and TTM tasks, we use Fréchet Audio Distance (FAD)~\cite{fad}, Fréchet Distance (FD), KL divergence, and CLAPScore~\cite{clap} as evaluation metrics. For the TTS task, we use Word Error Rate (WER), Speaker Similarity, and UTMOS~\cite{utmos} for evaluation.

\begin{table*}[t]
    \centering
    \caption{Reconstruction performance of discrete and continuous audio tokenizers. }
    \resizebox{0.9\textwidth}{!}{
    \begin{tabular}{l c c cc cc cc cc}
    \toprule
    \multirow{2}{*}{\textbf{Model}} 
    & \multirow{2}{*}{\textbf{\makecell{Frame\\Rate}}}
    & \multirow{2}{*}{\textbf{RTF}($\downarrow$)} 
    & \multicolumn{2}{c}{\textbf{AudioSet}}
    & \multicolumn{2}{c}{\textbf{MUSDB18}}
    & \multicolumn{2}{c}{\textbf{SeedTTS-EN}}
    & \multicolumn{2}{c}{\textbf{SeedTTS-ZH}} \\
    \cmidrule(lr){4-5} \cmidrule(lr){6-7} \cmidrule(lr){8-9} \cmidrule(lr){10-11}
    & & 
    & Mel-16k($\downarrow$) & STFT-16k($\downarrow$) 
    & Mel-16k($\downarrow$) & STFT-16k($\downarrow$) 
    & PESQ($\uparrow$) & STOI($\uparrow$) 
    & PESQ($\uparrow$) & STOI($\uparrow$) \\
    \midrule
    DAC          
        & 50 & 0.0043   
        & 0.615 & 1.884 & 0.578 & 1.755 & 3.786 & 0.969 & 3.877 & 0.967 \\
    SNAC         
        & -- & \textbf{0.0019}   
        & 1.156 & 2.700 & 1.172 & 2.600 & 1.817 & 0.872 & 1.879 & 0.867 \\
    XY-Tokenizer 
        & 12.5 & 0.0099 
        & 1.096 & 2.795 & 1.138 & 2.714 & 2.173 & 0.901 & 2.319 & 0.909 \\
    GLM-4-Voice
        & 12.5 & 0.0754 
        & -- & -- & -- & -- & 1.110 & 0.600 & 1.110 &  0.610 \\
    X-Codec 2
        & 50 & --
        & 1.263 & 3.000 & 1.275 & 2.852 & 2.086 & 0.897 & 2.250 & 0.901  \\

    \midrule
    EzAudio     
        & 50 & 0.0054 
        & \uline{0.298} & \uline{1.226} & \uline{0.259} & \uline{1.168} 
        & 3.649 & \uline{0.989} & 3.850 & 0.987 \\
    UniFlow-Audio 
        & 50 & 0.0167 
        & \textbf{0.268} & \textbf{1.185} & \textbf{0.232} & \textbf{1.129} 
        & 3.833 & \textbf{0.992} & 4.040 & \textbf{0.991} \\
    MingTok-Audio
        & 50 & 0.0050 
        & 0.500 & 1.949 & 0.472 & 1.792 
        & \uline{3.976} & 0.983 & \textbf{4.199} & 0.983 \\
    DashengTokenizer
        & 25 & 0.0034 
        & 0.370 & 1.736 & 0.329 & 1.584 
        & \textbf{4.122} & 0.987 & \uline{4.164} & \uline{0.988} \\
    \hdashline
    \textbf{LoSATok}
        & 25 & \uline{0.0033} 
        & 0.760 & 2.296 & 0.712 & 2.159 
        & 3.051 & 0.947 & 3.157 & 0.947 \\
    
    \bottomrule
    \end{tabular}
    }
    \label{tab:reconstruction}
\end{table*}

\noindent\textbf{Results.}
Table \ref{tab:generation} demonstrates our strong performance on cross-domain generation tasks. Compared with the acoustic tokenizer from UniFlow-Audio, our semantically rich representation is better suited for DiT learning under the same parameter settings, leading to superior generation quality. For example, when trained only on TTS, LoSATok achieves SIM and UTMOS scores of 0.548 and 3.367, respectively, exceeding UniFlow-Audio’s 0.408 and 2.768. For the high-dimensional unified representation DashengTokenizer, LoSATok achieves clearly better performance despite using fewer trainable parameters (208M vs. 322M). For the 1280-dimensional representation, substantially more parameters are required to achieve performance comparable to LoSATok (208M vs. 975M), and DashengTokenizer performs poorly on the TTS task. Moreover, LoSATok achieves better or competitive single-task performance with substantially fewer trainable parameters, and is especially strong on TTS and TTA semantic alignment. 
Moreover, downstream DiT models achieve the lowest inference cost with LoSATok.
These results demonstrate the benefits of our low-dimensional unified representation for cross-domain generation.

\subsection{Impact of DiT Dimensions}
\noindent\textbf{Dimension Setup.} 
Prior studies show the hidden dimension of DiT also affects generation performance~\cite{rae}. We therefore aim to investigate how different DiT dimensions influence acoustic and unified tokenizers. Specifically, we set the DiT dimension to 128, 512, and 1024, respectively, while keeping the number of DiT layers fixed at 12 and all other settings unchanged. 

\begin{figure}[t]
    \centerline{
    \includegraphics[width=0.4\textwidth]{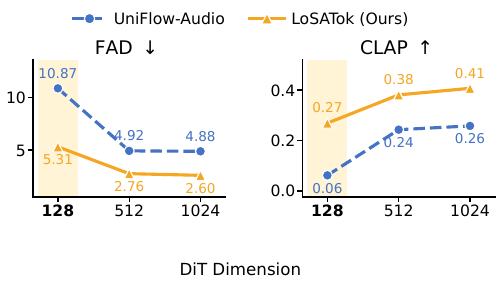}}
    \caption{DiT dimensions for downstream generation.}
    \label{fig:exp_dim}
\end{figure}

\noindent\textbf{Results.}
From Fig.~\ref{fig:exp_dim}, we find that when the DiT dimension is 128, i.e., equal to the representation dimension, the acoustic representation has almost no generation capability. For example, its CLAP score is only 0.06, while the FAD reaches 10.87, indicating a large distribution gap between the generated and real audio and almost no semantic alignment with the text prompt. In contrast, LoSATok still maintains non-trivial generation capability at 128 dimensions, achieving performance comparable to UniFlow-Audio with DiT dimensions of 512 or 1024. 
This experiment provides intuitive evidence of the significant difference between the semantically rich LoSATok and acoustic tokenizers across different generation settings.

\subsection{Audio Reconstruction}

\noindent\textbf{Baselines.} 
For discrete NACs, we compare against DAC~\cite{dac}, SNAC~\cite{snac}, XY-Tokenizer~\cite{xy-tokenizer}, GLM-4-Voice-Tokenizer~\cite{glm-4-voice}, and X-Codec 2~\cite{llasa}. 
For CATs, we utilize audio tokenizers from EzAudio~\cite{ezaudio}, UniFlow-Audio~\cite{uniflow-audio}, MingTok-Audio~\cite{ming-uniaudio}, and DashengTok~\cite{dashengtokenizer}.

\noindent\textbf{Evaluation.} 
We measure the Mel-spectrogram (Mel-16k) and Short-Time Fourier Transform (STFT-16k) distances on MUSDB18~\cite{musdb18} and AudioSet-eval~\cite{audioset}. We use the wide-band Perceptual Evaluation of Speech Quality (PESQ) and Short-Time Objective Intelligibility (STOI) as evaluation metrics on SeedTTS (EN/ZH)~\cite{seedtts}.

\noindent\textbf{Results.} 
Table \ref{tab:reconstruction} shows the reconstruction performance. We find that although LoSATok outperforms some NACs, e.g., SNAC and XY-Tokenizer, in reconstruction quality, it still lags behind advanced acoustic tokenizers. For a representation designed for generation and understanding, reconstruction metrics mainly reflect the degree of distortion. The reconstruction score may not be directly related to generation. For example, UniFlow-Audio achieves the best reconstruction performance, but its understanding and generation capabilities are inferior to those of LoSATok. We also conduct subjective evaluations on the generated audio in Appendix \ref{sec:subjective}, and the results suggest LoSATok-based generation has good perceptual quality.

\begin{figure}[t]
    \centerline{
    \includegraphics[width=0.4\textwidth]{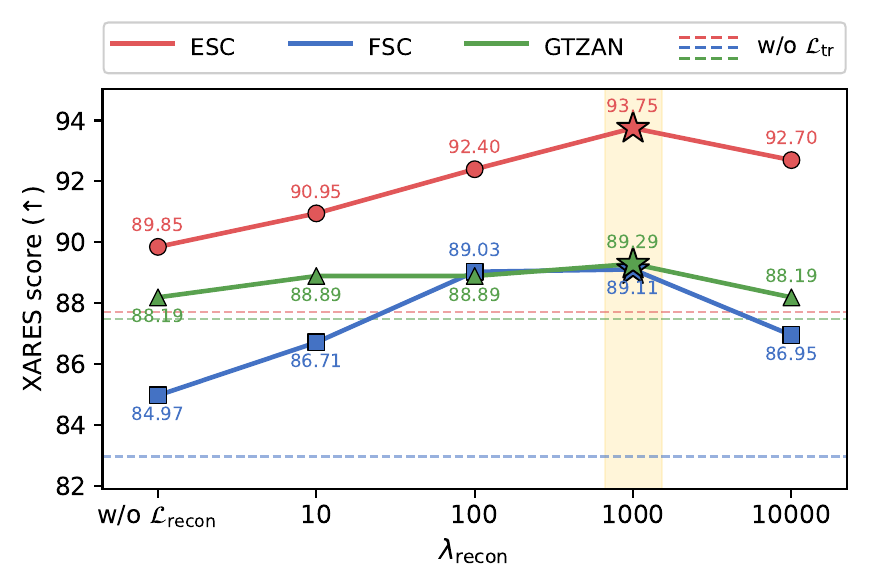}}
    \caption{Ablation results of Semantic Bottleneck.}
    \label{fig:ablation_sembo}
\end{figure}

\subsection{Ablation Study}\label{sec:exp_ablations}
\noindent\textbf{Components of Semantic Bottleneck.}
In this section, we evaluate the effects of $\mathcal{L}_{\mathrm{recon}}$ and $\mathcal{L}_{\mathrm{tr}}$ in SemBo, as well as the sensitivity of $\lambda_{\mathrm{recon}}$. Fig. \ref{fig:ablation_sembo} shows the results of the ablation study. We find that removing $\mathcal{L}_{\mathrm{tr}}$ leads to a large performance drop on the ESC and FSC understanding tasks, with scores of 87.70 and 82.97, respectively. Removing $\mathcal{L}_{\mathrm{recon}}$ also degrades understanding performance, but the results are slightly better than those obtained without (``w/o'') $\mathcal{L}_{\mathrm{tr}}$, which verifies the effectiveness of our proposed time-relation loss. Combining $\mathcal{L}_{\mathrm{recon}}$ and $\mathcal{L}_{\mathrm{tr}}$ yields better semantic performance. We further vary $\lambda_{\mathrm{recon}}$ from $10$ to $10^4$ on the understanding tasks and find that when $\lambda_{\mathrm{recon}} = 10^3$, the model achieves the best average cross-domain understanding performance, with scores of 93.75 on ESC, 89.11 on FSC, and 89.29 on GTZAN. Therefore, we set $\lambda_{\mathrm{recon}} = 10^3$ in Eq. (\ref{eq:sembo}).

\noindent\textbf{Components of LoSATok.}
We explore the effectiveness of LoSATok's objectives. All models are trained with a global batch size of 16 for 100K steps. We exclude the KL divergence to remove its potential influence on the results. Thus, this section uses an autoencoder (AE). KL-divergence experiments are provided in Appendix \ref{sec:kl}. Table \ref{tab:ablation_losatok} reports the results. We find that both $\mathcal{L}_{\mathrm{H}}$ and $\mathcal{L}_{\mathrm{L}}$ in LoSATok's dual-level semantic supervision affect understanding tasks. Removing $\mathcal{L}_{\mathrm{L}}$ almost eliminates the representation's understanding ability, while removing $\mathcal{L}_{\mathrm{H}}$ also leads to a certain performance drop on the FSC task. Although Section \ref{sec:sembo} shows that a training-free approach can also preserve high-dimensional understanding ability, this experiment shows that the training-free CM strategy is unsuitable as a semantic supervision signal, as it causes a substantial drop in understanding performance. These results validate the effectiveness of SemBo and LoSATok's design.

\begin{table}[t]
    \centering
    \caption{Ablation results of the LoSATok.}
    \label{tab:ablation_losatok}
    \resizebox{0.5\textwidth}{!}{
    \begin{tabular}{lccccccccc}
    \toprule
    \multirow{3}{*}{\textbf{Method}}
    & \multicolumn{6}{c}{\textbf{Reconstruction}}
    & \multicolumn{3}{c}{\textbf{Understanding}} \\
    \cmidrule(lr){2-7} \cmidrule(lr){8-10}
    & \multicolumn{2}{c}{\textbf{AudioSet}}
    & \multicolumn{2}{c}{\textbf{MUSDB}}
    & \multicolumn{2}{c}{\textbf{SeedTTS-EN}}
    & \multirow{2}{*}{\textbf{ESC}}
    & \multirow{2}{*}{\textbf{FSC}}
    & \multirow{2}{*}{\textbf{GTZAN}} \\
    \cmidrule(lr){2-3} \cmidrule(lr){4-5} \cmidrule(lr){6-7}
    & Mel-16k & STFT-16k
    & Mel-16k & STFT-16k
    & PESQ & STOI
    & & & \\
    \midrule
    w/o $\mathcal{L}_{\mathrm{H}}$
    & 0.759 & 2.315
    & 0.700 & 2.161
    & 2.952 & 0.944
    & 91.10 & 54.79 & \textbf{86.99} \\
    
    w/o $\mathcal{L}_{\mathrm{L}}$
    & 0.585 & 2.111
    & 0.558 & 1.998
    & 3.121 & 0.953 
    & 47.25 & 6.30 & 53.76 \\
    
    w/ CM
    & \textbf{0.575} & \textbf{2.076}
    & \textbf{0.538} & \textbf{1.955}
    & \textbf{3.178} & \textbf{0.956}
    & 52.45 & 5.11 & 56.26 \\
    
    
    AE
    & 0.776 & 2.340
    & 0.705 & 2.167
    & 2.909 & 0.942
    & \textbf{91.25} & \textbf{59.87} & 86.49 \\
    \bottomrule
    \end{tabular}
    }
\end{table}

\noindent\textbf{Projection-Adapter Baseline.} A possible alternative to learning an explicitly supervised low-dimensional latent is to compress an existing high-dimensional tokenizer before DiT modeling. We therefore freeze DashengTokenizer and train a $1280\rightarrow128\rightarrow1280$ linear adapter for 200K steps on the same data used by LoSATok. The downstream DiT is trained on the 128-dimensional bottleneck using the setup in Section \ref{sec:exp_generation}. As shown in Table \ref{tab:adapter}, the adapter retains considerably more reconstruction fidelity than semantic accessibility: STOI remains 0.952, but ESC-50 decreases from 96.40 to 37.05, and GTZAN from 89.99 to 54.96. Its TTS performance is also consistently below LoSATok in three metrics, e.g., SIM 0.476 vs. 0.548. These results show that post-hoc dimensionality reduction does not preserve the organization of the semantic space; explicit low-dimensional semantic supervision is necessary.

\begin{table}[t]
    \centering
    \caption{Comparison of LoSATok with a linear projection on DashengTokenizer.}
    \label{tab:comparison}
    \resizebox{0.5\textwidth}{!}{
    \begin{tabular}{lc cc ccc ccc}
    \toprule
    \multirow{2}{*}{Model}
    & \multirow{2}{*}{\textbf{\makecell{Latent \\ Dim.}}}
    & \multirow{2}{*}{\textbf{STOI}}
    & \multicolumn{2}{c}{\textbf{Understanding}}
    & \multicolumn{3}{c}{\textbf{TTS}} \\
    \cmidrule(lr){4-5}
    \cmidrule(lr){6-8}
    & &
    & ESC-50 & GTZAN
    & WER$\downarrow$ & SIM$\uparrow$ & UTMOS$\uparrow$ \\
    \midrule
    
    DashengTok
    & 1280
    & \textcolor{gray}{0.987}
    & \textcolor{gray}{96.40}
    & \textcolor{gray}{89.99}
    & \textcolor{gray}{100.0}
    & \textcolor{gray}{0.015}
    & \textcolor{gray}{1.251} \\
    
    \multicolumn{1}{r}{+ adapter}
    & 128 & \textbf{0.952}
    & 37.05 & 54.96
    & 3.456 & 0.476 & 2.623 \\
    
    \addlinespace[3pt]
    
    \textbf{LoSATok}
    & 128 & 0.947
    & \textbf{88.90} & \textbf{85.39}
    & \textbf{3.030} & \textbf{0.548} & \textbf{3.367} \\
    
    \bottomrule
    \end{tabular}
    }
    \label{tab:adapter}
\end{table}

\section{Conclusion}
In this paper, we propose LoSATok, a unified low-dimensional tokenizer for cross-domain understanding and generation.
We introduce low-dimensional semantic representations based on time-relation loss and dual-level semantic supervision to enrich semantics.
On understanding tasks, LoSATok matches several SSL models. On generation tasks, LoSATok outperforms acoustic and high-dimensional unified audio tokenizers.

\section*{Acknowledgment}
This work is supported by National Natural Science Foundation of China (62076144) and Shenzhen Science and Technology Program (JCYJ20220818101014030).

\section*{Limitations}
LoSATok aims to propose a tokenizer for semantic-acoustic modeling that can support both understanding and generation tasks. Its advantage lies in introducing a low-dimensional unified representation that is easier for DiT to learn, while still preserving useful semantic and acoustic information for speech, music, and general audio tasks.

However, LoSATok is not intended to replace tokenizers designed for high-fidelity reconstruction. Compared with acoustic tokenizers, LoSATok sacrifices a certain degree of reconstruction fidelity in order to obtain a low-dimensional representation that has a stronger semantic structure and is more suitable for generative modeling. 
Nevertheless, both subjective and objective experiments demonstrate its advantages in generation tasks.

In addition, although LoSATok retains a certain level of audio understanding capability and outperforms some SSL representations, e.g., HuBERT and WavLM, its semantic ability has not yet fully reached or approached the level of high-dimensional semantic representations, e.g., MiDashengLM. In our framework, SemBo is mainly used to demonstrate that high-dimensional semantic representations can be effectively compressed, while LoSATok introduces acoustic information and KL regularization to support downstream generation tasks. This design involves trade-offs among semantics, acoustics, and generation. Further optimizing the balance among these three aspects within a low-dimensional representation is a promising direction for future research.

\normalem
\bibliography{reference}

@article{dashengtokenizer,
  title={DashengTokenizer: One layer is enough for unified audio understanding and generation},
  author={Dinkel, Heinrich and Sun, Xingwei and Li, Gang and Mei, Jiahao and Niu, Yadong and Liu, Jizhong and Li, Xiyang and Liao, Yifan and Zhou, Jiahao and Zhang, Junbo and others},
  journal={arXiv preprint arXiv:2602.23765},
  year={2026}
}

@article{wavcube,
  title={WavCube: Unifying Speech Representation for Understanding and Generation via Semantic-Acoustic Joint Modeling},
  author={Yang, Guanrou and Tan, Tian and Chen, Qian and Niu, Zhikang and Song, Yakun and Ma, Ziyang and Chen, Yushen and Xie, Zeyu and Wang, Tianrui and Yang, Yifan and others},
  journal={arXiv preprint arXiv:2605.06407},
  year={2026}
}

@article{midashenglm,
  title={Midashenglm: Efficient audio understanding with general audio captions},
  author={Dinkel, Heinrich and Li, Gang and Liu, Jizhong and Luan, Jian and Niu, Yadong and Sun, Xingwei and Wang, Tianzi and Xiao, Qiyang and Zhang, Junbo and Zhou, Jiahao},
  journal={arXiv preprint arXiv:2508.03983},
  year={2025}
}

@inproceedings{xares,
  title     = {{X-ARES: A Comprehensive Framework for Assessing Audio Encoder Performance}},
  author    = {Junbo Zhang and Heinrich Dinkel and Yadong Niu and Chenyu Liu and Si Cheng and Anbei Zhao and Jian Luan},
  year      = {2025},
  booktitle = {{Interspeech 2025}},
  pages     = {4868--4872},
  doi       = {10.21437/Interspeech.2025-552},
  issn      = {2958-1796},
}

@article{dac,
  title={High-fidelity audio compression with improved rvqgan},
  author={Kumar, Rithesh and Seetharaman, Prem and Luebs, Alejandro and Kumar, Ishaan and Kumar, Kundan},
  journal={Advances in Neural Information Processing Systems},
  volume={36},
  pages={27980--27993},
  year={2023}
}

@inproceedings{rae,
    title={Diffusion Transformers with Representation Autoencoders},
    author={Boyang Zheng and Nanye Ma and Shengbang Tong and Saining Xie},
    booktitle={The Fourteenth International Conference on Learning Representations (ICLR)},
    year={2026},
}

@article{ming-uniaudio,
  title={Ming-UniAudio: Speech LLM for Joint Understanding, Generation and Editing with Unified Representation},
  author={Yan, Canxiang and Jin, Chunxiang and Huang, Dawei and Yu, Haibing and Peng, Han and Zhan, Hui and Gao, Jie and Peng, Jing and Chen, Jingdong and Zhou, Jun and others},
  journal={arXiv preprint arXiv:2511.05516},
  year={2025}
}

@article{uniflow-audio,
  title={Uniflow-audio: Unified flow matching for audio generation from omni-modalities},
  author={Xu, Xuenan and Mei, Jiahao and Zheng, Zihao and Tao, Ye and Xie, Zeyu and Zhang, Yaoyun and Liu, Haohe and Wu, Yuning and Yan, Ming and Wu, Wen and others},
  journal={arXiv preprint arXiv:2509.24391},
  year={2025}
}

@article{xy-tokenizer,
  title={Xy-tokenizer: Mitigating the semantic-acoustic conflict in low-bitrate speech codecs},
  author={Gong, Yitian and Jin, Luozhijie and Deng, Ruifan and Zhang, Dong and Zhang, Xin and Cheng, Qinyuan and Fei, Zhaoye and Li, Shimin and Qiu, Xipeng},
  journal={arXiv preprint arXiv:2506.23325},
  year={2025}
}

@inproceedings{audioset,
  title={Audio set: An ontology and human-labeled dataset for audio events},
  author={Gemmeke, Jort F and Ellis, Daniel PW and Freedman, Dylan and Jansen, Aren and Lawrence, Wade and Moore, R Channing and Plakal, Manoj and Ritter, Marvin},
  booktitle={2017 IEEE international conference on acoustics, speech and signal processing (ICASSP)},
  pages={776--780},
  year={2017},
  organization={IEEE}
}

@inproceedings{ualm,
    title={{UALM}: Unified Audio Language Model for Understanding, Generation and Reasoning},
    author={Jinchuan Tian and Sang-gil Lee and Zhifeng Kong and Sreyan Ghosh and Arushi Goel and Chao-Han Huck Yang and Wenliang Dai and Zihan Liu and Hanrong Ye and Shinji Watanabe and Mohammad Shoeybi and Bryan Catanzaro and Rafael Valle and Wei Ping},
    booktitle={The Fourteenth International Conference on Learning Representations (ICLR)},
    year={2026}
}

@inproceedings{dit,
  title={Scalable diffusion models with transformers},
  author={Peebles, William and Xie, Saining},
  booktitle={Proceedings of the IEEE/CVF international conference on computer vision},
  pages={4195--4205},
  year={2023}
}

@article{style-transfer,
  title={Neural style transfer: A review},
  author={Jing, Yongcheng and Yang, Yezhou and Feng, Zunlei and Ye, Jingwen and Yu, Yizhou and Song, Mingli},
  journal={IEEE transactions on visualization and computer graphics},
  volume={26},
  number={11},
  pages={3365--3385},
  year={2019},
  publisher={IEEE}
}

@inproceedings{voxcpm,
    title={Hierarchical Semantic-Acoustic Modeling via Semi-Discrete Residual Representations for Expressive End-to-End Speech Synthesis},
    author={Yixuan Zhou and Guoyang Zeng and Xin Liu and Xiang Li and Renjie Yu and Ziyang Wang and Runchuan Ye and Weiyue Sun and Jiancheng Gui and Kehan Li and Zhiyong Wu and Zhiyuan Liu},
    booktitle={The Fourteenth International Conference on Learning Representations (ICLR)},
    year={2026},
}

@inproceedings{levo,
    title={LeVo: High-Quality Song Generation with Multi-Preference Alignment},
    author={Shun Lei and Yaoxun Xu and Zhiwei Lin and Huaicheng Zhang and Wei Tan and Hangting Chen and Yixuan Zhang and Chenyu Yang and Haina Zhu and Shuai Wang and Zhiyong Wu and Dong Yu},
    booktitle={The Thirty-ninth Annual Conference on Neural Information Processing Systems},
    year={2025},
}

@inproceedings{snac,
    title={{SNAC}: Multi-Scale Neural Audio Codec},
    author={Hubert Siuzdak and Florian Gr{\"o}tschla and Luca A Lanzend{\"o}rfer},
    booktitle={Audio Imagination: NeurIPS 2024 Workshop AI-Driven Speech, Music, and Sound Generation},
    year={2024}
}

@article{llasa,
  title={Llasa: Scaling train-time and inference-time compute for llama-based speech synthesis},
  author={Ye, Zhen and Zhu, Xinfa and Chan, Chi-Min and Wang, Xinsheng and Tan, Xu and Lei, Jiahe and Peng, Yi and Liu, Haohe and Jin, Yizhu and Dai, Zheqi and others},
  journal={arXiv preprint arXiv:2502.04128},
  year={2025}
}

@article{semanticvocoder,
  title={SemanticVocoder: Bridging Audio Generation and Audio Understanding via Semantic Latents},
  author={Xie, Zeyu and Li, Chenxing and Jin, Qiao and Xu, Xuenan and Yang, Guanrou and Wang, Wenfu and Wu, Mengyue and Yu, Dong and Zou, Yuexian},
  journal={arXiv preprint arXiv:2602.23333},
  year={2026}
}

@inproceedings{vocos,
    title={Vocos: Closing the gap between time-domain and Fourier-based neural vocoders for high-quality audio synthesis},
    author={Hubert Siuzdak},
    booktitle={The Twelfth International Conference on Learning Representations (ICLR)},
    year={2024},
}

@inproceedings{librispeech,
  title={Librispeech: an asr corpus based on public domain audio books},
  author={Panayotov, Vassil and Chen, Guoguo and Povey, Daniel and Khudanpur, Sanjeev},
  booktitle={2015 IEEE international conference on acoustics, speech and signal processing (ICASSP)},
  pages={5206--5210},
  year={2015},
  organization={IEEE}
}

@article{vctk,
  title={CSTR VCTK Corpus: English multi-speaker corpus for CSTR voice cloning toolkit (version 0.92)},
  author={Yamagishi, Junichi and Veaux, Christophe and MacDonald, Kirsten},
  journal={The Rainbow Passage which the speakers read out can be found in the International Dialects of English Archive:(http://web. ku. edu/\~{} idea/readings/rainbow. htm).},
  year={2019},
  publisher={University of Edinburgh. The Centre for Speech Technology Research (CSTR)}
}

@inproceedings{jamendo,
  title={The mtg-jamendo dataset for automatic music tagging},
  author={Bogdanov, Dmitry and Won, Minz and Tovstogan, Philip and Porter, Alastair and Serra, Xavier},
  booktitle={Machine learning for music discovery workshop, international conference on machine learning (ICML 2019)},
  pages={1--3},
  year={2019},
  organization={Long Beach, CA, United States}
}

@inproceedings{musdb18,
  title={MUSDB18 - a corpus for music separation},
  author={Zafar Rafii and Antoine Liutkus and Fabian-Robert Stöter and Stylianos Ioannis Mimilakis and Rachel Bittner},
  year={2017},
  url={https://api.semanticscholar.org/CorpusID:199578935}
}

@inproceedings{ezaudio,
  title     = {{EzAudio: Enhancing Text-to-Audio Generation with Efficient Diffusion Transformer}},
  author    = {Jiarui Hai and Yong Xu and Hao Zhang and Chenxing Li and Helin Wang and Mounya Elhilali and Dong Yu},
  year      = {2025},
  booktitle = {{Interspeech 2025}},
  pages     = {4233--4237},
  issn      = {2958-1796},
}

@inproceedings{common_voice,
  title={Common voice: A massively-multilingual speech corpus},
  author={Ardila, Rosana and Branson, Megan and Davis, Kelly and Kohler, Michael and Meyer, Josh and Henretty, Michael and Morais, Reuben and Saunders, Lindsay and Tyers, Francis and Weber, Gregor},
  booktitle={Proceedings of the twelfth language resources and evaluation conference},
  pages={4218--4222},
  year={2020}
}

@article{seedtts,
  title={Seed-tts: A family of high-quality versatile speech generation models},
  author={Anastassiou, Philip and Chen, Jiawei and Chen, Jitong and Chen, Yuanzhe and Chen, Zhuo and Chen, Ziyi and Cong, Jian and Deng, Lelai and Ding, Chuang and Gao, Lu and others},
  journal={arXiv preprint arXiv:2406.02430},
  year={2024}
}

@inproceedings{audiocaps,
  title={Audiocaps: Generating captions for audios in the wild},
  author={Kim, Chris Dongjoo and Kim, Byeongchang and Lee, Hyunmin and Kim, Gunhee},
  booktitle={Proceedings of the 2019 Conference of the North American Chapter of the Association for Computational Linguistics: Human Language Technologies, Volume 1 (Long and Short Papers)},
  pages={119--132},
  year={2019}
}

@article{wavcaps,
  title={Wavcaps: A chatgpt-assisted weakly-labelled audio captioning dataset for audio-language multimodal research},
  author={Mei, Xinhao and Meng, Chutong and Liu, Haohe and Kong, Qiuqiang and Ko, Tom and Zhao, Chengqi and Plumbley, Mark D and Zou, Yuexian and Wang, Wenwu},
  journal={IEEE/ACM Transactions on Audio, Speech, and Language Processing},
  volume={32},
  pages={3339--3354},
  year={2024},
  publisher={IEEE}
}

@article{musiclm,
  title={Musiclm: Generating music from text},
  author={Agostinelli, Andrea and Denk, Timo I and Borsos, Zal{\'a}n and Engel, Jesse and Verzetti, Mauro and Caillon, Antoine and Huang, Qingqing and Jansen, Aren and Roberts, Adam and Tagliasacchi, Marco and others},
  journal={arXiv preprint arXiv:2301.11325},
  year={2023}
}

@article{libritts,
  title={Libritts: A corpus derived from librispeech for text-to-speech},
  author={Zen, Heiga and Dang, Viet and Clark, Rob and Zhang, Yu and Weiss, Ron J and Jia, Ye and Chen, Zhifeng and Wu, Yonghui},
  journal={arXiv preprint arXiv:1904.02882},
  year={2019}
}

@inproceedings{fad,
  title     = {{Fréchet Audio Distance: A Reference-Free Metric for Evaluating Music Enhancement Algorithms}},
  author    = {Kevin Kilgour and Mauricio Zuluaga and Dominik Roblek and Matthew Sharifi},
  year      = {2019},
  booktitle = {{Interspeech 2019}},
  pages     = {2350--2354},
  doi       = {10.21437/Interspeech.2019-2219},
  issn      = {2958-1796},
}

@inproceedings{clap,
  title={Clap learning audio concepts from natural language supervision},
  author={Elizalde, Benjamin and Deshmukh, Soham and Al Ismail, Mahmoud and Wang, Huaming},
  booktitle={ICASSP 2023-2023 IEEE International Conference on Acoustics, Speech and Signal Processing (ICASSP)},
  pages={1--5},
  year={2023},
  organization={IEEE}
}

@inproceedings{utmos,
  title     = {{UTMOS: UTokyo-SaruLab System for VoiceMOS Challenge 2022}},
  author    = {Takaaki Saeki and Detai Xin and Wataru Nakata and Tomoki Koriyama and Shinnosuke Takamichi and Hiroshi Saruwatari},
  year      = {2022},
  booktitle = {{Interspeech 2022}},
  pages     = {4521--4525},
  doi       = {10.21437/Interspeech.2022-439},
  issn      = {2958-1796},
}

@article{jmas-vae,
  title={On the Distillation Loss Functions of Speech VAE for Unified Reconstruction, Understanding, and Generation},
  author={Cheng, Changhao and Wang, Wei and Zhang, Wangyou and Jia, Dongya and Wu, Jian and Chen, Zhuo and Qian, Yanmin},
  journal={arXiv preprint arXiv:2604.12383},
  year={2026}
}

@article{semantic-vae,
  title={Semantic-VAE: Semantic-Alignment Latent Representation for Better Speech Synthesis},
  author={Niu, Zhikang and Hu, Shujie and Choi, Jeongsoo and Chen, Yushen and Chen, Peining and Zhu, Pengcheng and Yang, Yunting and Zhang, Bowen and Zhao, Jian and Wang, Chunhui and others},
  journal={arXiv preprint arXiv:2509.22167},
  year={2025}
}

@article{hubert,
  title={Hubert: Self-supervised speech representation learning by masked prediction of hidden units},
  author={Hsu, Wei-Ning and Bolte, Benjamin and Tsai, Yao-Hung Hubert and Lakhotia, Kushal and Salakhutdinov, Ruslan and Mohamed, Abdelrahman},
  journal={IEEE/ACM transactions on audio, speech, and language processing},
  volume={29},
  pages={3451--3460},
  year={2021},
  publisher={IEEE}
}

@article{wavlm,
  title={Wavlm: Large-scale self-supervised pre-training for full stack speech processing},
  author={Chen, Sanyuan and Wang, Chengyi and Chen, Zhengyang and Wu, Yu and Liu, Shujie and Chen, Zhuo and Li, Jinyu and Kanda, Naoyuki and Yoshioka, Takuya and Xiao, Xiong and others},
  journal={IEEE Journal of Selected Topics in Signal Processing},
  volume={16},
  number={6},
  pages={1505--1518},
  year={2022},
  publisher={IEEE}
}

@inproceedings{whisper,
  title={Robust speech recognition via large-scale weak supervision},
  author={Radford, Alec and Kim, Jong Wook and Xu, Tao and Brockman, Greg and McLeavey, Christine and Sutskever, Ilya},
  booktitle={International conference on machine learning},
  pages={28492--28518},
  year={2023},
  organization={PMLR}
}

@article{lp-musiccaps,
  title={Lp-musiccaps: Llm-based pseudo music captioning},
  author={Doh, SeungHeon and Choi, Keunwoo and Lee, Jongpil and Nam, Juhan},
  journal={arXiv preprint arXiv:2307.16372},
  year={2023}
}

@article{unisrcodec,
  title={UniSRCodec: Unified and Low-Bitrate Single Codebook Codec with Sub-Band Reconstruction},
  author={Zhang, Zhisheng and Li, Xiang and Zhou, Yixuan and Peng, Jing and Cai, Shengbo and Zeng, Guoyang and Wu, Zhiyong},
  journal={arXiv preprint arXiv:2601.02776},
  year={2026}
}

@article{qwen3-omni,
  title={Qwen3-omni technical report},
  author={Xu, Jin and Guo, Zhifang and Hu, Hangrui and Chu, Yunfei and Wang, Xiong and He, Jinzheng and Wang, Yuxuan and Shi, Xian and He, Ting and Zhu, Xinfa and others},
  journal={arXiv preprint arXiv:2509.17765},
  year={2025}
}

@article{glm-4-voice,
  title={Glm-4-voice: Towards intelligent and human-like end-to-end spoken chatbot},
  author={Zeng, Aohan and Du, Zhengxiao and Liu, Mingdao and Wang, Kedong and Jiang, Shengmin and Zhao, Lei and Dong, Yuxiao and Tang, Jie},
  journal={arXiv preprint arXiv:2412.02612},
  year={2024}
}

\appendix

\newpage
\section{Discussion}
In this section, we discuss some important aspects.

\noindent\textbf{Ethical Considerations.}
To evaluate the subjective perceptual quality of the generated audio, we perform a subjective listening test in Appendix \ref{sec:subjective}. These experiments have received approval from the local Human Ethics Research Committee. Before the experiment, we obtain informed consent from all participants. We do not collect any personal information, and all responses are used solely for academic research and kept strictly confidential. After completing the test, each participant receives approximately \$1.47 as compensation.

\noindent\textbf{Reconstruction and Generation.}
Prior studies~\cite{semanticvocoder} have found that although models do not perform best on reconstruction tasks, they can achieve strong performance on downstream generation tasks. From a representation perspective, high-quality reconstruction requires representations with sufficient acoustic details, whereas generation tasks require representations that are easier for downstream models, such as DiTs, to learn. In the XARES benchmark, these tasks are typically evaluated by directly feeding the representations into the MLP or LLM. LoSATok’s unified representations achieve comparable performance on understanding tasks. Compared with acoustically rich representations, semantically rich representations are easier for downstream models to learn, which explains why our model performs better in generation. Reconstruction evaluation mainly measures whether the tokenizer introduces audio distortion. The subjective experiments in Appendix \ref{sec:subjective} further show that audio generated based on LoSATok has good perceptual quality and outperforms acoustic tokenizers in generation tasks.
Overall, LoSATok is not designed to be the best reconstruction tokenizer. It sacrifices part of reconstruction fidelity to obtain a lower-dimensional, more semantically organized latent space that is easier for text-conditioned DiT generation.

\noindent\textbf{Term Explanation.}
In this paper, we use ``unified'' to refer to the integration of acoustic and semantic representations~\cite{dashengtokenizer} for both generation and understanding. We use ``cross-domain/general-purpose'' to refer to joint modeling across the three domains of speech, music, and general audio (sound). The paper uses different terms accordingly.

\begin{table*}[t]
    \centering
    \caption{Generation, understanding, and reconstruction performance under different KL loss weights.}
    \label{tab:kl_ablation}
    \resizebox{0.95\textwidth}{!}{
    \begin{tabular}{lccccccccccc}
    \toprule
    \multirow{3}{*}{$\lambda_{\mathrm{KL}}$} 
    & \multicolumn{3}{c}{\textbf{Generation}} 
    & \multicolumn{2}{c}{\textbf{Understanding}} 
    & \multicolumn{6}{c}{\textbf{Reconstruction}} \\
    \cmidrule(lr){2-4} \cmidrule(lr){5-6} \cmidrule(lr){7-12}
    & \multicolumn{3}{c}{\textbf{TTS}} 
    & \multirow{2}{*}{\textbf{ESC}$\uparrow$} 
    & \multirow{2}{*}{\textbf{GTZAN}$\uparrow$} 
    & \multicolumn{2}{c}{\textbf{AudioSet}} 
    & \multicolumn{2}{c}{\textbf{MUSDB}} 
    & \multicolumn{2}{c}{\textbf{SeedTTS-EN}} \\
    & WER$\downarrow$ & SIM$\uparrow$ & UTMOS$\uparrow$
    &  &  
    & Mel-16k$\downarrow$ & STFT-16k$\downarrow$ 
    & Mel-16k$\downarrow$ & STFT-16k$\downarrow$ 
    & PESQ$\uparrow$ & STOI$\uparrow$ \\
    \midrule
    w/o $\mathcal{L}_{\mathrm{KL}}$ 
    & 3.338 & 0.463 & 3.170 
    & \textbf{91.40} & 86.99 
    & \textbf{0.688} & \textbf{2.195} 
    & \textbf{0.626} & \textbf{2.027} 
    & \textbf{3.447} & \textbf{0.964} \\
    
    $10^{-4}$
    & 3.395 & 0.449 & 3.330 
    & 90.35 & 86.29 
    & 0.694 & 2.203 
    & 0.633 & 2.043 
    & 3.424 & 0.963 \\
    
    $10^{-3}$
    & 3.158 & 0.491 & 3.284 
    & 91.10 & \textbf{88.39} 
    & 0.694 & 2.203 
    & 0.636 & 2.044 
    & 3.405 & 0.962 \\
    
    $10^{-2}$
    & \textbf{3.030} & \textbf{0.548} & \textbf{3.367}
    & 88.90 & 85.39 
    & 0.760 & 2.296 
    & 0.712 & 2.159 
    & 3.051 & 0.947 \\
    \bottomrule
    \end{tabular}
    }
\end{table*}

\noindent\textbf{Downstream Fairness.}
To ensure fairness in evaluating generation tasks, we follow the multi-task training framework of UniFlow-Audio and replace only the VAE component, while keeping all other architectural components unchanged. We also keep the number of training GPUs, batch size, and training steps consistent, freeze the respective audio tokenizers, and train the DiT blocks and some connection layers. For understanding tasks, we feed the reparameterized representations from LoSATok into the XARES benchmark for evaluation.

\begin{table}[t]
    \centering
    \caption{Dimension reduction via FC layers.}
    \label{tab:dim_reduction_linear}
    \resizebox{0.4\textwidth}{!}{
    \begin{tabular}{lcccc}
    \toprule
    \textbf{Method} 
    & \textbf{Dim.} 
    & \textbf{ESC}($\uparrow$) 
    & \textbf{FSC}($\uparrow$) 
    & \textbf{GTZAN}($\uparrow$) \\
    \midrule
    SemBo
        & 128 & \textbf{93.75} & \textbf{89.11} & \textbf{89.29} \\
    FC-Semantic
        & 128 & 41.15 & 13.60 & 67.87 \\
    FC-Unified 
        & 128 & 42.55 & 4.75 & 56.96 \\
    \bottomrule
    \end{tabular}
    }
\end{table}

\noindent\textbf{Why Are Low-dimensional Semantic Representations Important?}
We train SemBo separately as the low-dimensional semantic signals. This is because, without an explicit low-dimensional semantic supervision target, learning semantic dimensionality reduction jointly with tokenizer reconstruction can cause the low-dimensional representation to be dominated by the acoustic reconstruction objective, making it difficult to preserve semantics.

To verify this, we construct a direct-projection baseline using DashengTokenizer. This baseline uses two trainable fully connected (FC) layers to reduce the dimensionality of the high-dimensional semantic and acoustic representations, respectively, and retains dual-level semantic supervision to obtain the low-dimensional unified representation. Specifically, we replace the low-dimensional semantic representation from SemBo with the semantic representation directly projected by the fully connected layer, so that semantic dimensionality reduction is optimized jointly with tokenizer reconstruction. The training architecture and implementation details are kept consistent with Section~\ref{sec:exp_ablations}.

We further evaluate understanding capabilities of the resulting low-dimensional unified representation on the XARES benchmark in Table \ref{tab:dim_reduction_linear}.
Compared with SemBo, semantic representations obtained via FC-based dimensionality reduction have significantly weaker comprehension ability. Consequently, the unified representations learned from such weak semantic supervision are also much worse than those from LoSATok. This indicates that applying dimensionality reduction through an FC layer together with reconstruction training, without using directly frozen semantic signals as teacher supervision, causes the trainable parameters to focus more on acoustic details and leads to severe loss of semantic information. In contrast, independently trained SemBo can provide more reliable low-dimensional semantic supervision, making it better suited to guide the learning of low-dimensional unified representations.

\section{KL Divergence in LoSATok}\label{sec:kl}
LoSATok improves downstream generation by utilizing the KL divergence. Prior studies have shown that KL divergence can regularize the representation distribution by mapping it to a more structured latent space~\cite{semantic-vae}, e.g., a Gaussian distribution, which facilitates modeling by downstream DiT models and further improves generation quality. However, a strong KL constraint may over-smooth the representations~\cite{wavcube}, while a weak KL constraint may provide limited benefits for generation. Therefore, in this section, we analyze the impact of KL divergence and its weight $\lambda_{\mathrm{KL}}$ on LoSATok. All experiments follow the same training setup as LoSATok, using a global batch size of 64 and training for one million steps.

We evaluate the model on generation, understanding, and reconstruction tasks under four settings: without the KL loss and reparameterization layers, and with KL weights of $10^{-4}$, $10^{-3}$, and $10^{-2}$. The results are shown in Table \ref{tab:kl_ablation}. When $\lambda_{\mathrm{KL}}=10^{-2}$, the model achieves the best performance on the TTS generation task, greatly outperforming both the setting without KL divergence and the setting with $\lambda_{\mathrm{KL}}=10^{-4}$. Specifically, WER decreases to 3.030\%, SIM increases to 0.548, and UTMOS increases to 3.367, indicating improvements in intelligibility, speaker similarity, and perceptual audio quality. Meanwhile, performance on the ESC and GTZAN understanding tasks does not degrade substantially. Although reconstruction performance under this setting drops notably compared with the AE architecture without KL divergence, we select $\lambda_{\mathrm{KL}}=10^{-2}$ as the default configuration due to its benefits for generation.

We further analyze the effects of different KL weights across tasks. KL divergence affects both understanding and reconstruction performance. The AE architecture without KL divergence performs better on the ESC understanding task, but its SIM score on TTS generation is only 0.463, which is relatively low. When the KL weight is too small, e.g., $\lambda_{\mathrm{KL}}=10^{-4}$, overall performance decreases across understanding, generation, and reconstruction tasks. In contrast, $\lambda_{\mathrm{KL}}=10^{-3}$ achieves a more balanced trade-off among the three types of tasks. To enhance generation performance without substantially compromising understanding performance, we adopt $\lambda_{\mathrm{KL}}=10^{-2}$ as the main experimental setting. This choice increases the SIM score of the TTS task from 0.491 to 0.548 compared with $\lambda_{\mathrm{KL}}=10^{-3}$, and further improves audio quality and intelligibility. To support different use cases, we plan to release the pre-trained weights for both $\lambda_{\mathrm{KL}}=10^{-2}$ and $\lambda_{\mathrm{KL}}=10^{-3}$.

\section{Effect of Data Source and Scale}

\begin{table}[t]
    \centering
    \caption{Data scaling experiments on TTA.}
    \resizebox{0.45\textwidth}{!}{
    \begin{tabular}{lcccccc}
    \toprule
    \multirow{2}{*}{\textbf{Model}} 
    & \multirow{2}{*}{\textbf{\# Param}} 
    & \multicolumn{4}{c}{\textbf{Text-to-Audio}} \\
    \cmidrule(lr){3-6}
     & 
     & FAD($\downarrow$) & FD($\downarrow$)
     & KL($\downarrow$) & CLAP($\uparrow$) \\
    \midrule
    \multicolumn{6}{l}{\textit{Large-scale Dataset, WavCaps, with $\sim$ 7,558 hours.}} \\
    UniFlow-Audio
        & 208M
        & 4.925 & 40.017 & 2.613 & 0.243 \\
    \textbf{LoSATok}
        & 208M
        & \textbf{2.760} & \textbf{25.743} & \textbf{1.844} & \textbf{0.381} \\
    \midrule
    \multicolumn{6}{l}{\textit{Small-scale Dataset, AudioCaps, with $\sim$ 123 hours.}} \\
    UniFlow-Audio
        & 208M
        & 2.425 & 23.101 & 1.688 & 0.428 \\
    \textbf{LoSATok}
        & 208M
        & \textbf{1.813} & \textbf{15.931} & \textbf{1.273} & \textbf{0.507} \\
    \bottomrule
    \end{tabular}
    }
    \label{tab:data_scale}
\end{table}

We aim to evaluate the performance of acoustic and unified tokenizers under different training data scales. In this section, we conduct experiments on the large-scale WavCaps dataset ($\sim$7,558 hours) and the small-scale AudioCaps dataset ($\sim$123 hours). Both models are trained for 100K steps on 4 NVIDIA H100 GPUs.

Table \ref{tab:data_scale} presents the TTA results under different data scales. On the small-scale AudioCaps dataset, the acoustic tokenizer also achieves competitive performance, with an FAD of 2.425 and a CLAP score of 0.428. This indicates that the distribution of the generated audio is close to that of real audio and that the generated content follows the text prompts reasonably well. However, on the large-scale WavCaps dataset, the limitations of the acoustic tokenizer become more evident. Its FAD and CLAP scores are only 4.925 and 0.243, respectively, indicating degraded overall audio quality and text-audio alignment. In contrast, LoSATok achieves stable and superior performance across both data scales. Specifically, it obtains an FAD of 1.813 and a CLAP score of 0.507 when trained on AudioCaps, and an FAD of 2.760 and a CLAP score of 0.381 when trained on WavCaps.

This experiment highlights the gap between tokenizers that rely solely on acoustic information and those with rich acoustic and semantic representations in generation tasks. On large-scale datasets, LoSATok converges faster and achieves better generation quality than the acoustic tokenizer. On small-scale datasets, although both tokenizers perform well, LoSATok still demonstrates a stronger downstream generation advantage.

\section{Human Study}\label{sec:subjective}

\begin{figure}[t]
    \centerline{
    \includegraphics[width=0.5\textwidth]{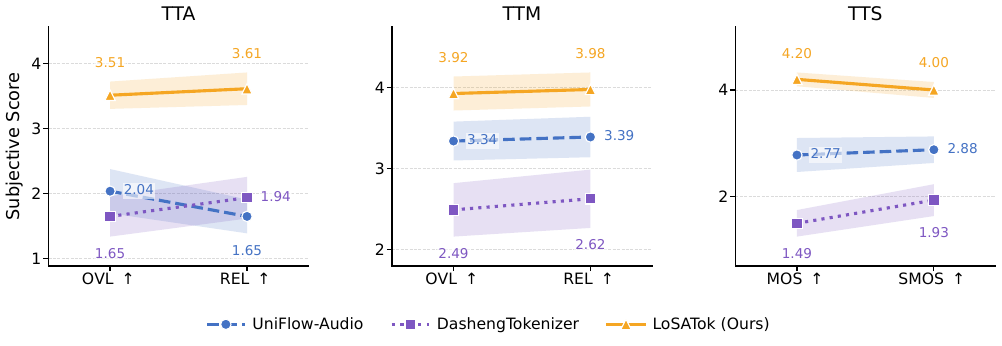}}
    \caption{Subjective results on generation tasks with 95\% confidence intervals.}
    \label{fig:subjective}
\end{figure}

We test the objective metrics of the generation task in Section \ref{sec:exp_generation}. In this section, we evaluate the subjective perception of the generated audio.

\noindent\textbf{Participant Profile.} We design a subjective evaluation questionnaire on the Credamo platform and recruit 20 participants for the experiment. All participants are over 18 years old, and their informed consent is obtained. Their average response time is 397.15 seconds. Each participant receives approximately \$1.47 as compensation.

\noindent\textbf{Metrics and Questionnaire Design.} We design the subjective experiment following UniFlow-Audio~\cite{uniflow-audio}. For the TTA and TTM tasks, we evaluate the generated audio in terms of overall audio quality (OVL) and text relevance (REL). OVL measures the naturalness, clarity, and overall listening quality of the generated audio, while REL measures its consistency with the text description. For the TTS task, we evaluate speech naturalness (MOS) and speaker similarity (SMOS). All metrics are rated on a 0–5 scale, where 0 indicates very poor quality or a complete mismatch, and 5 indicates excellent quality or a high degree of matching. Participants are instructed to wear headphones in a quiet environment and rate each sample only after listening to the entire audio. 

\noindent\textbf{Results and Analysis.} Fig.~\ref{fig:subjective} presents the results of our subjective evaluation. We report the mean scores and 95\% confidence intervals (CI), shown as the shaded regions in the figure. For the TTA and TTM tasks, LoSATok outperforms UniFlow-Audio and DashengTokenizer in both OVL and REL, with large improvements on TTA. Specifically, LoSATok achieves a REL score of $3.61 \pm 0.25$, while UniFlow-Audio and DashengTokenizer achieve only $1.65 \pm 0.26$ and $1.94 \pm 0.32$, respectively. For the TTS task, LoSATok achieves a MOS score of $4.20 \pm 0.13$, outperforming the baselines. These results demonstrate that LoSATok achieves better perceptual quality and text relevance in generation tasks. Although LoSATok does not achieve the best reconstruction performance, its unified representation is better suited for downstream DiT-based generation tasks.

\section{Detailed Information}

\subsection{Details of Effective Rank and PCA Analysis}\label{sec:details_rank}
In Section~\ref{sec:sembo}, we analyze the effective rank and PCA spectra of the semantic and acoustic representations of DashengTokenizer to study the compressibility of high-dimensional semantic representations. In this section, we provide a detailed explanation of the relevant concepts.

Given a frozen semantic encoder $\mathrm{E}_s$, it outputs a semantic latent $z_s^h = \mathrm{E}_s(x) \in \mathbb{R}^{T \times D}$, where $T$ is the number of temporal frames and $D = 1280$ is the representation dimension. We collect frame-level representations over the dataset and compute the covariance matrix along the feature dimension.

First, we compute the effective rank of the high-dimensional semantic representations. 
Let $\lambda_1 \geq \lambda_2 \geq \cdots \geq \lambda_D$ be the eigenvalues of the feature covariance matrix $\Sigma\in\mathbb{R}^{D\times D}$ computed across all frame-level representations in the dataset. We normalize them into a probability distribution:
\begin{equation}
    p_i = \frac{\lambda_i}{\sum_{j=1}^{D}\lambda_j}.
\end{equation}

The effective rank is then defined in Eq. (\ref{eq:eff_rank}).
\begin{equation}
    r_{\mathrm{eff}} = \exp\left(-\sum_{i=1}^{D} p_i \log p_i\right) \in [1, D].
    \label{eq:eff_rank}
\end{equation}
The effective rank measures the number of dimensions over which the feature energy is effectively distributed.

Second, we perform PCA analysis and compute the number of principal components required to preserve a fraction $\alpha$ of the total variance:
\begin{equation}
    k_{\alpha} = \min \left\{k:
    \frac{\sum_{i=1}^{k}\lambda_i}
    {\sum_{j=1}^{D}\lambda_j} \geq \alpha \right\}.
\end{equation}

In our analysis, we set $\alpha=0.9$.

Fig.~\ref{fig:dim} shows that the effective dimensionality of the 1280-dimensional semantic representations can be substantially lower than their original dimensionality, indicating that high-dimensional semantic representations contain considerable redundancy and are therefore potentially compressible.

\subsection{Details of SemBo}
The SemBo architecture consists of a two-layer MLP-based compressor and restorer, each with 0.72M parameters. For the semantic encoder, we use the audio encoder of MiDashengLM~\cite{midashenglm}, considering its strong performance on the audio understanding leaderboard on the XARES benchmark. We train SemBo for 100K steps using the same 13K-hour training dataset as LoSATok. The training details are summarized in Table~\ref{tab:details_sembo}. We use AdamW as the optimizer, with $\beta_1$ and $\beta_2$ set to 0.8 and 0.99, respectively. The initial learning rate is $1 \times 10^{-4}$, and a cosine learning-rate scheduler is adopted with 1,000 warmup steps.

\begin{table}[t]
    \centering
    \caption{Training hyperparameters for SemBo.}
    \resizebox{0.25\textwidth}{!}{
    \label{tab:sembo_training}
    \begin{tabular}{ll}
    \toprule
    \textbf{Parameter} & \textbf{Value} \\
    \midrule
    optimizer & AdamW \\
    $[\beta_1, \beta_2]$ & $[0.8, 0.99]$ \\
    lr & $1 \times 10^{-4}$ \\
    lr\_scheduler & CosineLR \\
    warmup\_steps & 1,000 \\
    max\_steps & 100,000 \\
    \bottomrule
    \end{tabular}
    }
    \label{tab:details_sembo}
\end{table}

\subsection{Details of LoSATok}\label{sec:details_losatok}
\noindent\textbf{LoSATok Architecture.} 
LoSATok is designed based on DashengTokenizer. The encoder consists of a semantic encoder and an acoustic encoder, with detailed parameter information summarized in Table~\ref{tab:details_losatok}. The semantic encoder is composed of a pretrained MiDashengLM encoder followed by SemBo, and its output dimension is 128. The acoustic encoder applies a 2D convolutional layer to perform non-overlapping compression, and then uses a linear layer to project the feature dimension to 128. Two linear layers are used to compute the KL divergence based on the unified embedding $z_{\mathrm{uni}}$, enabling adaptation to downstream DiT-based generation tasks. The acoustic decoder is based on Vocos~\cite{vocos} and reconstructs audio details from the unified representation.

For the 25Hz tokenizer, given an input audio signal $x$, we extract its Mel-spectrogram $x_{\mathrm{mel}} \in \mathbb{R}^{T, F}$, where $F$ denotes the number of Mel bins. The acoustic encoder uses 2D convolutions to compress the frequency dimension to 1 and downsample the time dimension by a factor of 4, producing a 25Hz representation. Since the semantic representation is also at 25Hz, it can be directly added to the generated representation to obtain a unified representation. In the decoder, it is first upsampled to 50 Hz, then passed through Vocos blocks, followed by an Inverse Short-Time Fourier Transform (ISTFT) head to reconstruct the audio.

\begin{table}[t]
    \centering
    \caption{Architecture details of LoSATok.}
    \label{tab:losatok_architecture}
    \resizebox{0.45\textwidth}{!}{
    \begin{tabular}{lclc}
    \toprule
    Module & Parameters & Architecture & Dim. \\
    \midrule
    Semantic Encoder & 632 M & \makecell[l]{32-layer Transformer \\ + SemBo} & 128 \\
    Acoustic Encoder & 0.82 M & 2D Conv + Linear & 128 \\
    KL divergence & 0.03 M & 2 $\times$ Linear & 128 \\
    Decoder & 171 M & Vocos & 128 \\
    \bottomrule
    \end{tabular}
    }
    \label{tab:details_losatok}
\end{table}

\noindent\textbf{Details of LoSATok's Objectives.} We describe LoSATok’s training objectives in Eq. (\ref{eq:losatok}).

\begin{itemize}[leftmargin=1em, topsep=0pt, itemsep=0pt, parsep=0pt]
    \item $\mathcal{L}_{\mathrm{mel}}$: The multi-scale Mel-spectrogram $L_1$ loss. We compare the reconstructed waveform $\hat{x}$ with the original waveform $x$ using multiple window sizes $[32, 64, \ldots, 2048]$ and different numbers of Mel bins $[5, 10, \ldots, 320]$, thereby ensuring the perceptual quality of the reconstructed audio.

    \item $\mathcal{L}_{\mathrm{sem}} = \mathcal{L}_{\mathrm{H}} + \mathcal{L}_{\mathrm{L}}$: The semantic alignment loss. We perform $\ell_2$ alignment between acoustic representations and their corresponding high- and low-dimensional semantic targets, thereby improving the understanding capability of the general-purpose representation.

    \item $\mathcal{L}_{\mathrm{KL}}$: The KL divergence loss. At the low-dimensional bottleneck, we compute the closed-form KL divergence between the posterior distribution
    $q(z \mid x) = \mathcal{N}(\mu, \sigma^2 I)$
    and the prior distribution $\mathcal{N}(0, I)$:
    \begin{equation}
        \mathcal{L}_{\mathrm{KL}} =
        -\frac{1}{2} \sum_d
        \left(1 + \log \sigma_d^2 - \mu_d^2 - \sigma_d^2\right),
    \end{equation}
    where $\mu$ denotes the mean and $\sigma^2$ denotes the variance.
    The unified representation is then reparameterized and fed into the decoder.

    \item $\mathcal{L}_{\mathrm{fm}}$: The feature matching loss based on a multi-frequency discriminator. Given real and generated audio as inputs, we apply an $L_1$ matching loss to each intermediate feature, which helps stabilize adversarial training.

    \item $\mathcal{L}_{\mathrm{adv}}$: The hinge-style adversarial loss, which improves the generation quality through the discriminator.
\end{itemize}

\noindent\textbf{LoSATok Training Details.} LoSATok uses the same training dataset as SemBo in Table \ref{tab:datasets}. Compared to SemBo in Table \ref{tab:details_sembo}, the number of training steps is increased to one million, and the minimum learning rate is $1 \times 10^{-5}$ for sufficient training.

\subsection{Details of Training Datasets}
To train SemBo and LoSATok, we select approximately 13K hours of training data, as shown in Table \ref{tab:datasets}. To achieve general-purpose capabilities, we keep the proportions of different domains roughly balanced. In addition, to evaluate the generation capability of LoSATok on downstream tasks, we follow previous studies \cite{uniflow-audio, dashengtokenizer} and select multiple datasets. For the LP-MusicCaps dataset, because MusicCaps is used as the test set, we only use the MagnaTagATune (MTT) subset for training. The joint multi-task training in Section \ref{sec:exp_generation} uses LibriTTS, LP-MusicCaps, and WavCaps as large-scale training datasets. AudioCaps is used for comparison in the data scaling experiments.

\begin{table}[t]
    \centering
    \caption{Datasets for LoSATok training and generation.}
    \resizebox{0.4\textwidth}{!}{
    \begin{tabular}{clcc}
    \toprule
    \textbf{Domain} & \textbf{Dataset} & \textbf{Subset} & \textbf{Hours (h)} \\
    \midrule
    \multicolumn{4}{l}{\textit{Dataset for LoSATok}} \\
    \multirow{3}{*}{Speech}
    & VCTK & train & 82 \\
    & Common Voice & train (EN) & 3,507 \\
    & LibriSpeech & train & 961 \\
    \midrule
    \multirow{2}{*}{Music}
    & MTG--Jamendo & train & 3,763 \\
    & MUSDB & train & 3 \\
    \midrule
    General Audio
    & AudioSet & bal \& unbal & 4,836 \\
    \midrule
    \multicolumn{4}{l}{\textit{Dataset for Generation}} \\
    Speech
    & LibriTTS & train & 523 \\
    \midrule
    Music
    & LP-MusicCaps & MTT--train & 125 \\
    \midrule
    \multirow{2}{*}{General Audio}
    & WavCaps & train & 7,558 \\
    & AudioCaps & train & 123 \\
    \bottomrule
    \end{tabular}
    }
    \label{tab:datasets}
\end{table}

\begin{figure*}[t]
    \centerline{
    \includegraphics[width=0.95\textwidth]{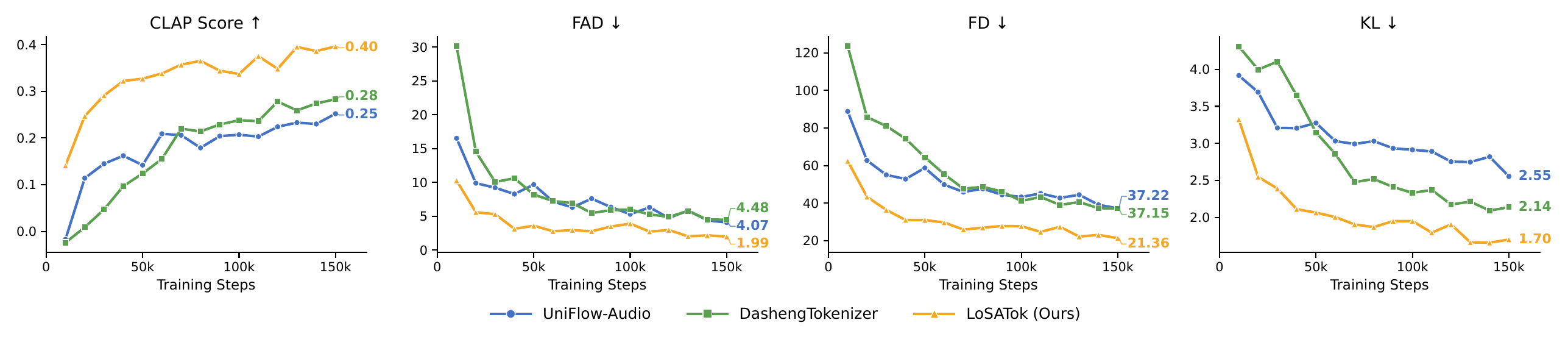}}
    \caption{Text-to-audio training process of different audio tokenizers.}
    \label{fig:generation_curve}
\end{figure*}

\subsection{Details of Downstream Generation}


\noindent\textbf{Training-Curve Visualization.}
Figure \ref{fig:generation_curve} shows the comparison curves of LoSATok, UniFlow-Audio, and DashengTokenizer in terms of convergence speed and performance during multi-task joint training, with evaluation conducted every 10K steps. The experimental results show that LoSATok converges faster. In terms of the CLAP metric, LoSATok reaches performance comparable to that achieved by UniFlow-Audio and DashengTokenizer at 150K training steps after only 21K and 28K steps, respectively, saving approximately 120K training steps. This result verifies the effectiveness of LoSATok in improving training convergence efficiency.

\noindent\textbf{Evaluation Metrics.}
For different generation tasks, we follow \cite{uniflow-audio} and adopt the following evaluation strategies:

\begin{itemize}[leftmargin=1em, topsep=0pt, itemsep=0pt, parsep=0pt]
    \item TTA \& TTM: For the FD metric, we compute the distributional distance between generated and reference audio based on features extracted by PANNs CNN14. The CLAP metric is used to measure the semantic alignment between the input prompt text and the generated audio, and we conduct the evaluation using the pretrained weights~\footnote{\href{https://github.com/LAION-AI/CLAP}{https://github.com/LAION-AI/CLAP}} released by LAION-AI.
    \item TTS: For the WER metric, we adopt a pretrained model~\footnote{\href{https://huggingface.co/nvidia/stt_en_conformer_transducer_xlarge}{https://huggingface.co/nvidia/stt\_en\_conformer\_transducer\_xlarge}} from the NeMo toolkit to evaluate the intelligibility of generated speech. For the SIM metric, we use pretrained weights based on WavLM-large fine-tuned on the speaker verification task~\footnote{\href{https://github.com/BytedanceSpeech/seed-tts-eval}{https://github.com/BytedanceSpeech/seed-tts-eval}} to measure the speaker similarity between synthesized and real speech. For the UTMOS metric, we use it as an objective measure of the perceptual quality of synthesized audio.
\end{itemize}

\subsection{Real Time Factor}
In Table~\ref{tab:reconstruction}, we compute the real-time factor (RTF) to evaluate the efficiency of audio tokenizers. RTF is defined as the total time required by the tokenizer to reconstruct an audio sample divided by the duration of the audio. For each model, we conduct the evaluation on a single NVIDIA H100 GPU, with no additional GPU processes running during testing. We measure the RTF on the four datasets listed in Table~\ref{tab:reconstruction} and report the average value.

The results show that LoSATok achieves an RTF of only 0.0033, outperforming several CATs, e.g., UniFlow-Audio, and is therefore able to meet practical real-time requirements.

\end{document}